\documentclass[a4paper,fleqn]{cas-sc}
\usepackage[numbers]{natbib}
\usepackage{booktabs}
\usepackage{adjustbox}
\usepackage{longtable}
\usepackage{caption}
\usepackage{mathtools}

\def\tsc#1{\csdef{#1}{\textsc{\lowercase{#1}}\xspace}}
\tsc{WGM}
\tsc{QE}
\tsc{EP}
\tsc{PMS}
\tsc{BEC}
\tsc{DE}

\begin{document}
\let\WriteBookmarks\relax
\def\floatpagepagefraction{1}
\def\textpagefraction{.001}
\let\printorcid\relax

\shorttitle{Accepted for publication in Applied Mathematics and Computation}

\shortauthors{Chaoqian Wang {\it et~al.}}

\title [mode = title]{Inertia in spatial public goods games under weak selection}                      

\author[1]{Chaoqian Wang}
\ead{CqWang814921147@outlook.com}
\credit{xxx}

\author[2]{Attila Szolnoki}
\cormark[1]
\cortext[cor1]{Corresponding author}
\ead{szolnoki.attila@ek-cer.hu}
\credit{xx}

\address[1]{Department of Computational and Data Sciences, George Mason University, Fairfax, VA 22030, USA}
\address[2]{Institute of Technical Physics and Materials Science, Centre for Energy Research, P.O. Box 49, H-1525 Budapest, Hungary}

\begin{abstract}
Due to limited cognitive skills for perceptual error or other emotional reasons, players may keep their current strategies even if there is a more promising choice. Such behavior inertia has already been studied, but its consequences remained unexplored in the weak selection limit. To fill this gap, we consider a spatial public goods game model where inertia is considered during the imitation process. By using the identity-by-descent method, we present analytical forms of the critical synergy factor $r^\star$, which determines when cooperation is favored. We find that inertia hinders cooperation, which can be explained by the decelerated coarsening process under weak selection. Interestingly, the critical synergy conditions for different updating protocols, including death-birth and birth-death rules, can be formally linked by the extreme limits of the inertia factor. To explore the robustness of our observations, calculations are made for different lattices and group sizes. Monte Carlo simulations also confirm the results.
\end{abstract}


\begin{highlights}
\item The impact of inertia is studied in public goods games in the weak selection limit
\item Analytical results are derived for the critical synergy factor of cooperation success
\item Inertia links the cooperation success conditions for different updating rules
\item Theoretical results are confirmed by Monte Carlo simulations
\end{highlights}


\maketitle

\section{Introduction}\label{intro}
According to the evolutionary game principle, a strategy with a larger fitness is more successful and eventually conquers the whole system~\cite{maynard_82}. However, the final outcome may depend on several circumstances besides the key parameters that characterize the actual social dilemma where cooperation and defection fight~\cite{sigmund_10,szabo_pr07,roca_plr09}. For example, it could be an essential factor whether players carrying competing strategies are well-mixed or they are fixed in a way that their interactions can be well described by a graph~\cite{szabo1998evolutionary}. In the former case, which is frequently called as a structured population, a specific cooperator-supporting mechanism may emerge~\cite{nowak1992evolutionary}. This is called as network reciprocity, which is based on the fact that cooperators may accumulate and support each other against the external invasion of defectors. Furthermore, it could also be a critical detail which type of microscopic dynamical rules is used. The latter choice may determine how a player's strategy varies during an elementary process. In particular, four major dynamical rules were studied intensively in the last decades, including the so-called birth-death, death-birth, imitation, and pairwise comparison strategy updating~\cite{ohtsuki_jtb06,allen2017evolutionary}. Nevertheless, the variety of potential models does not stop at this level  because it could also be a free choice how the payoff originated from interactions determines the fitness of a player~\cite{traulsen_jtb07b,fu_pre09b,zhou_l_pre18}. If this link is relevant, then we call it a strong or intermediate selection, while in the weak selection limit, the payoff has just a gentle contribution to fitness. In general, the latter scenario makes it possible to apply analytically feasible calculations~\cite{lieberman2005evolutionary,ohtsuki2006simple,ibsen2015computational}, while in the former case, our observations are mainly based on numerical efforts~\cite{liu_rr_amc19,wang2013interdependent,quan_j_c19,zhang_lm_pa21,li_k_csf21}. To explore the robustness of observations originating from different approaches, we must clarify the differences and possible overlaps of alternative predictions.

Thanks to the intensive and fruitful collaboration of the research community, we also identified further critical details in the last decades~\cite{liu_lj_rspa22,ohdaira_srep22,han_jrsif22}. They are frequently called mechanisms, which may have decisive impacts on the basic competition of cooperator and defector strategies~\cite{nowak2006five}. It would be a hopeless task to list them all in a single Introduction section. Instead, we here refer to topical reviews, most of which are discussed in detail~\cite{perc2017statistical,rand_tcs13,perc_jrsi13}. From our present perspective, however, it is particularly interesting when the microscopic dynamical rule is influenced by an effect that modifies the updating probability slightly. For example, this could result from a subjective viewpoint of how players see their neighborhood~\cite{amaral_rspa20,wang_z_pre14b,huang_k_pa18}. One can argue easily that our personal success is more valuable for us than the achievement of others, hence the comparison of fitness values, which practically determines the willingness to change a strategy, is less objective~\cite{li_k_srep16,szolnoki_pre18}. In other words, players may be reluctant to change their status, which can be termed a kind of inertia~\cite{szolnoki2009impact,liu2010effects,zhang2011inertia}. Alternatively, a conceptually similar effect can be reached if we introduce an additional cost of strategy change~\cite{szabo2002phase}. 

In the following, we briefly summarize the possible consequences of inertia on the evolution of cooperation. It is worth noting that inertia has different implications for selection strength. While selection strength measures the importance of payoff in strategy updating (how eager individuals learn the strategy with a higher payoff), inertia measures the reluctance of individuals to change strategies (whether the strategy brings a high payoff or not). The inertia effect seems to be a strategy-neutral intervention because it is not biased toward any competing strategies. Notably, the possible consequences were already studied by several previous papers. These works considered both well-mixed~\cite{zhang2011inertia} and structured populations~\cite{liu2010effects,du2012effects,chang2018cooperation} in strong selection. As we previously argued, this liberty offers a large variety of models, and quite interestingly, the case of analytical calculation in structured populations under weak selection remained largely unexplored. While the case of two-player games has been investigated previously~\cite{wang2023evolution}, the specific goal of our paper is to fill the gap for spatial public goods games. The importance of our work is not simply to complete the zoo of potential models but to check the robustness of previous observations. It was found, for example, that appropriate inertia can promote cooperation on various graphs~\cite{du2012effects,chang2018cooperation}. Furthermore, the diversity of inertia and other mechanisms also generates more subtle phenomena~\cite{liu2010effects,zhang2011inertia,jia2018effects,he2020behavior}. 

Our principal goal is to obtain analytical results by applying the so-called identity-by-descent (IBD) method~\cite{allen2014games,su2019spatial} and calculate the theoretical condition of cooperation success on transitive graphs in finite populations. According to the definition of transitive graphs, all nodes are indistinguishable in perceiving their positions on the graph by solely observing the neighborhood. In this way, all lattices with periodic boundary conditions are transitive and can be studied by this method. Although arbitrary two-player games have been solved on any population structure by further considering coalescing random walk method in IBD~\cite{allen2017evolutionary}, the accurate solutions for multiplayer games in finite populations were only explored on transitive graphs~\cite{su2019spatial,su2018understanding}. Earlier, theoretical solutions for multiplayer games were also calculated by pair approximation on random regular graphs in an infinite population~\cite{li2014cooperation,li2015evolutionary,li2016evolutionary}, but it is slightly different from the frequently studied spatial dilemma situation.
The simplest prototype of multiplayer games is the public goods game, where multi-point interactions are not necessarily a simple sum of pairwise interactions~\cite{perc2013evolutionary,battiston2020networks,burgio2020evolution,alvarez2021evolutionary}. In the original case, where the payoff is a linear function of the number of participating cooperators, the underlying public goods game may be interpreted as a superposition of the prisoner's dilemma game, which makes the solution explicit~\cite{su2019spatial}. But in general, multiplayer games in structured populations remain a complex and demanding challenge~\cite{li2016evolutionary}.

Besides analytical results, we also provide numerical calculations to check our findings. Additionally, we demonstrate that different values of the inertia parameter make possible the transition between alternative versions of the model, including death-birth, imitation, and birth-death dynamics. This way, we present analytical results for these types of updating rules.

\section{Model}\label{secmodel}
To model a spatially structured population, we consider an $L\times L$ square lattice with periodic boundary conditions. Each node is occupied by an agent, hence the population $N=L^2$. Each agent interacts with the $k$ nearest neighbors. On a square lattice, it means $k=4$ (von~Neumann neighborhood), which is the scope of our main study, but extensions to the larger neighborhood, including $k=8$ (Moore neighborhood), $k=12$, and $k=24$ are also discussed. Each agent $i$ forms a group $\Omega_i$ centered on itself, containing $i$'s neighbors and the focal $i$, hence $G=k+1$ agents in the group. As a result, each agent also belongs to $G$ groups, centered on itself and its $k$ neighbors, respectively. Because of the graph's transitiveness, every agent is involved in $G$ public goods games.

We denote the strategy of agent $i$ by $s_i$. The agent can employ either cooperation ($s_i=1$) or defection ($s_i=0$) at each elementary Monte Carlo (MC) step. During an elementary step, we randomly select a focal agent $i$ for a potential strategy update. In the public goods game centered on agent $g\in\Omega_i$, each player $j\in\Omega_g$ contributes $c$ ($c>0$) if cooperating, or contributes nothing if defecting. The accumulated contributions from cooperative players, $\sum_{j\in\Omega_g}s_j c$, is enlarged by a synergy factor $r$ ($r>1$) and redistributed to all $G$ players. Therefore, the income of agent $i$ from the group centered on agent $g$ is $r\sum_{j\in\Omega_g}s_j c/G-s_i c$. For the actual $\pi_i$ payoff of player $i$, we average the incomes collected from its $G$ related groups,
\begin{equation}\label{eqpayoff}
    \pi_i=\frac{1}{G}\sum_{g\in\Omega_i}\left(\frac{r\sum_{j\in\Omega_g}s_j c}{G}-s_i c\right).
\end{equation}

Next, we update the strategy of agent $i$ according to a modified imitation rule: the focal player $i$ keeps its current strategy or adopts one of the neighbors' strategies proportional to fitness. We calculate the payoff of $i$'s neighbors $\ell\in\Omega_i\backslash\{i\}$ in the same way introduced previously. Then, we transform the payoff values to fitness. Earlier works frequently assumed the fitness function $F_\ell=1+\pi_\ell/\kappa$~\cite{allen2017evolutionary,allen2014games}, where $\kappa$ is a noise parameter, whose inverse characterizes selection intensity.
This mapping is the first-order Taylor expansion of $\exp(\pi_\ell/\kappa)$ in the weak selection intensity $1/\kappa\to 0^+$. However, we do not see the necessity of approximation here and will take the complete form $F_\ell=\exp(\pi_\ell/\kappa)$. 

To describe the behavior of inertia, we introduce an additional parameter $\tau$. 
In the transformation of payoff to fitness, the focal agent $i$ adds this $\tau$ value beside the calculated payoff, hence the modified fitness $\exp(\tau+\pi_i/\kappa)$. In contrast, the fitness of non-focal neighbors is calculated in the original way. Therefore,
\begin{align}\label{eqfit}
    \begin{cases} 
    \displaystyle{F_i~=\exp(\tau+\pi_i/\kappa)}, & \mbox{when agent $i$ is the focal actor,}\\
    \displaystyle{F_\ell=\exp(\pi_\ell/\kappa)}, & \mbox{when agent $\ell\in\Omega_i\backslash\{i\}$ is a non-focal group member.}
    \end{cases}
\end{align}

According to the imitation protocol,
agent $i$ learns the strategy of an agent $j$ in the group $\Omega_i$ with the probability $W(s_i\gets s_j)$ proportional to $j$'s fitness,
\begin{equation}\label{eqimitation}
	W(s_i\gets s_j)
	=\frac{F_j}{\sum_{l\in \Omega_i}F_l}
	=
	\begin{cases} 
    \displaystyle{\frac{\exp(\tau+\pi_j/\kappa)}
	{\exp(\tau+\pi_i/\kappa)+\sum_{\ell\in \Omega_i\backslash\{i\}}\exp(\pi_\ell/\kappa)}},  & \mbox{if $j=i$,}\\
    \displaystyle{\frac{\exp(\pi_j/\kappa)}
	{\exp(\tau+\pi_i/\kappa)+\sum_{\ell\in \Omega_i\backslash\{i\}}\exp(\pi_\ell/\kappa)}}, & \mbox{if $j\neq i$.}
    \end{cases}
\end{equation}
Evidently, the learning probabilities are normalized, $\sum_{j\in \Omega_i} W(s_i\gets s_j)=1$. If $j=i$, agent $i$ keeps the original strategy. If $j\neq i$, agent $i$ must adopt the strategy of a neighbor. The above-described elementary MC step is repeated for $N$ times, which establishes a full MC step, so that all players have a chance to update the strategy on average.

At $\tau=0$, we turn back to the traditional imitation model. Intuitively, as $\tau$ increases, agents become more reluctant to change strategies. It is worth stressing that formally, the value of $\tau$ could also be negative, which has a specific meaning, as discussed later. In the following, our principal goal is to reveal the potential consequence of $\tau$ parameter on the evolution of cooperation in the weak selection limit.

\section{Theoretical analysis}\label{sectheo}
To calculate the critical value $r^\star$ of the synergy factor analytically, we adopt the IBD method~\cite{allen2014games}.
Cooperation is favored in the limit of weak selection strength if $r>r^\star$. We will define ``weak selection strength'' in Section~\ref{sectheoupdate}, and the meaning of ``cooperation success'' will be presented in Section~\ref{coop_suc}. 
But first, we reorganize the payoff calculation by using random walk formalism.

\subsection{Payoff calculation}
We define $n$-step random walks on the lattice; for example, an agent $i$'s $1$-step random walk ends in one of $i$'s neighbors. The expected payoff of agents with $n$ steps away is denoted by $\pi^{(n)}$. The probability of arriving at a cooperative agent after an $n$-step random walk is denoted by $s^{(n)}$.
These notations are unified among different agents $i$ by the transitiveness of the lattice.

Since self-loop is not allowed, we use the notation $k$ (the number of neighbors) to help clarify random walk steps. 
In the game organized by the $n$-step focal agent, there are $k$ co-players from $n+1$ steps away plus the $n$-step focal player. In the remaining $k$ games organized by the $n+1$-step neighbors, $k$ co-players are second-order neighbors from $n+2$ steps away plus the $n+1$-step focal player. Accordingly, payoff calculation can be written as
\begin{align}
	\pi^{(n)}&=\frac{1}{k+1}\left\{\left(\frac{r(k s^{(n+1)}+s^{(n)})c}{G}-s^{(n)}c\right)+k\left(\frac{r(k s^{(n+2)}+s^{(n+1)})c}{G}-s^{(n)}c\right)\right\}\nonumber
	\\
	&=\left(\frac{r}{(k+1)G}-1\right)s^{(n)}c+\frac{2k}{(k+1)G}rs^{(n+1)}c+\frac{k^2}{(k+1)G} rs^{(n+2)}c,
	\label{piwalk}
\end{align}
which will be used for calculation later.

\subsection{Updating strategies}\label{sectheoupdate}
In Eq.~(\ref{piwalk}), we can see a common factor $c$, which can be extracted and merged into the noise parameter $\kappa$ for the fitness term in Eq.~(\ref{eqfit}). Hence, we can introduce selection strength, denoted by $\delta=c/\kappa$. In this work, we consider weak selection strength $0<\delta\ll 1\Leftrightarrow 0<c\ll \kappa$.

Under neutral drift at $\delta=0$, the system can fix onto the full cooperation state with probability $N_C/N$ starting with $N_C$ cooperative agents~\cite{cox_ap83,nowak2004emergence}.
For simplicity, we check the initial state of one cooperator ($N_C=1$) in our theoretical analysis. In the following, we denote the initial single cooperator by player 1.

According to Refs.~\cite{allen2014games,nowak2010evolution}, the condition of cooperation success in the weak selection limit is
\begin{equation}\label{eqbd}
	\left\langle\frac{\partial}{\partial\delta}(\mathcal{B}_1-\mathcal{D}_1)\right\rangle_{\begin{smallmatrix}\delta=0\\s_1=1\end{smallmatrix}}>0,
\end{equation}
where $\langle\cdot\rangle$ means the expectation, $\mathcal{B}_1$ is the probability that player 1 reproduces its strategy, and $\mathcal{D}_1$ is the probability that player 1's strategy is replaced by the alternative strategy. When inertia is introduced into the transition probability of Eq.~(\ref{eqimitation}), the specific form of $\mathcal{B}_1$ and $\mathcal{D}_1$ are
\begin{subequations}\label{eqbdimi}
	\begin{align}
		\mathcal{B}_1&=\sum_{i\in \Omega_1\backslash\{1\}}\frac{1}{N}W(s_i\gets s_1)
		=\sum_{i\in \Omega_1\backslash\{1\}}\frac{1}{N}
		\frac{\exp(\pi_1/\kappa)}
	    {\exp(\tau+\pi_i/\kappa)+\sum_{\ell\in \Omega_i\backslash\{i\}}\exp(\pi_\ell/\kappa)},
	    \label{eqbdimib}
		\\
		\mathcal{D}_1&=\frac{1}{N}\sum_{j\in \Omega_1\backslash\{1\}}W(s_1\gets s_j)
		=\frac{1}{N}\sum_{j\in \Omega_1\backslash\{1\}}\frac{\exp(\pi_j/\kappa)}
	    {\exp(\tau+\pi_1/\kappa)+\sum_{\ell\in \Omega_1\backslash\{1\}}\exp(\pi_\ell/\kappa)}.
	    \label{eqbdimid}
	\end{align}
\end{subequations}
Player 1's strategy reproduces, when one of its neighbors, $i\in \Omega_1\backslash\{1\}$, is selected as the focal agent (with probability $1/N$) and learns 1's strategy with probability $W(s_i\gets s_1)$ given by
Eq.~(\ref{eqimitation}).
This constitutes the expression for $\mathcal{B}_1$ in Eq.~(\ref{eqbdimib}). Similarly, player 1's strategy is changed, when player 1 is selected as the focal agent (with probability $1/N$) and adopts the strategy of a neighbor $j\in \Omega_1\backslash\{1\}$ with probability $W(s_1\gets s_j)$ defined by Eq.~(\ref{eqimitation}).
This gives the expression for $\mathcal{D}_1$ in Eq.~(\ref{eqbdimid}).

By using the actual form of $\mathcal{B}_1$ and $\mathcal{D}_1$ in Eq.~(\ref{eqbdimi}), and noting the selection strength $\delta=c/\kappa$, we can calculate the condition of Eq.~(\ref{eqbd}),
\begin{align}\label{condigroup}
	&\left\langle\frac{\partial}{\partial\delta}(\mathcal{B}_1-\mathcal{D}_1)\right\rangle_{\begin{smallmatrix}\delta=0\\s_1=1\end{smallmatrix}}>0\nonumber
	\\
	\Leftrightarrow&~\frac{1}{N(k+\mathrm{e}^\tau)^2}
	\left(
	k(k+\mathrm{e}^\tau)\left\langle \pi_1\right\rangle_{\begin{smallmatrix}\delta=0\\s_1=1\end{smallmatrix}}
	-\mathrm{e}^\tau\left\langle \sum_{j\in \Omega_1\backslash\{1\}}\pi_j\right\rangle_{\begin{smallmatrix}\delta=0\\s_1=1\end{smallmatrix}}
	-\left\langle \sum_{i\in \Omega_1\backslash\{1\}}\sum_{\ell\in \Omega_i\backslash\{i\}}\pi_\ell\right\rangle_{\begin{smallmatrix}\delta=0\\s_1=1\end{smallmatrix}}
	\right)\nonumber
	\\
	&~-\frac{1}{N(k+\mathrm{e}^\tau)^2}
	\left(
	-k \mathrm{e}^\tau\left\langle \pi_1\right\rangle_{\begin{smallmatrix}\delta=0\\s_1=1\end{smallmatrix}}
	+\mathrm{e}^\tau\left\langle \sum_{j\in \Omega_1\backslash\{1\}}\pi_j\right\rangle_{\begin{smallmatrix}\delta=0\\s_1=1\end{smallmatrix}}
	\right)>0\nonumber
	\\
	\Leftrightarrow&\left\langle\pi_1\right\rangle_{\begin{smallmatrix}\delta=0\\s_1=1\end{smallmatrix}}
	-\frac{2\mathrm{e}^\tau}{k(k+2\mathrm{e}^\tau)}\left\langle \sum_{j\in \Omega_1\backslash\{1\}}\pi_j\right\rangle_{\begin{smallmatrix}\delta=0\\s_1=1\end{smallmatrix}}
	-\frac{1}{k(k+2\mathrm{e}^\tau)}\left\langle \sum_{i\in \Omega_1\backslash\{1\}}\sum_{\ell\in \Omega_i\backslash\{i\}}\pi_\ell\right\rangle_{\begin{smallmatrix}\delta=0\\s_1=1\end{smallmatrix}}>0.
\end{align}

Note that player 1 can be any agent selected by random in the population initially. Therefore, in the case where player 1 is the starting node of random walks, we have
\begin{equation}\label{eqtranstowalk}
    \pi^{(0)}=\left\langle\pi_1\right\rangle_{\begin{smallmatrix}\delta=0\\s_1=1\end{smallmatrix}},~
    \pi^{(1)}=\frac{1}{k}\left\langle \sum_{j\in \Omega_1\backslash\{1\}}\pi_j\right\rangle_{\begin{smallmatrix}\delta=0\\s_1=1\end{smallmatrix}},~
    \pi^{(2)}=\frac{1}{k^2}\left\langle \sum_{i\in \Omega_1\backslash\{1\}}\sum_{\ell\in \Omega_i\backslash\{i\}}\pi_\ell\right\rangle_{\begin{smallmatrix}\delta=0\\s_1=1\end{smallmatrix}}.
\end{equation}
In this way, Eq.~(\ref{condigroup}) can be simplified, hence the condition of cooperation success:
\begin{equation}\label{eqimicondi}
	\left\langle\frac{\partial}{\partial\delta}(\mathcal{B}_1-\mathcal{D}_1)\right\rangle_{\begin{smallmatrix}\delta=0\\s_1=1\end{smallmatrix}}>0
	\Leftrightarrow
	\pi^{(0)}
	-\frac{2\mathrm{e}^\tau}{k+2\mathrm{e}^\tau}\pi^{(1)}
	-\frac{k}{k+2\mathrm{e}^\tau}\pi^{(2)}>0.
\end{equation}

\subsection{The critical synergy factor for cooperation success}\label{coop_suc}
To apply the IBD method introduced in Ref.~\cite{allen2014games}, we temporarily assume strategy mutation with probability $\mu$, but later we will demonstrate that this parameter can be eliminated since $\mu\to 0$. According to Ref.~\cite{allen2014games}, in the low mutation limit $\mu\to 0$, we have
\begin{equation}\label{eqsub}
	s^{(n)}-s^{(n+1)}=\frac{\mu}{2}(Np^{(n)}-1)+\mathcal{O}(\mu^2),
\end{equation}
which transforms the calculation of strategies into the random walk behavior $p^{(n)}$. Here, $p^{(n)}$ means the probability of an $n$-step random walk ending in the starting node, which can be directly imagined according to specific network structures and will be demonstrated later. $\mathcal{O}(\mu^2)\to 0$ is a negligible term.

To achieve our goal, we need to transform Eq.~(\ref{eqsub}) as follows.
\begin{align}\label{eqsubtrans}
	&~s^{(n)}-\frac{2\mathrm{e}^\tau}{k+2\mathrm{e}^\tau}s^{(n+1)}-\frac{k}{k+2\mathrm{e}^\tau}s^{(n+2)}\nonumber\\
	=&~\frac{2\mathrm{e}^\tau}{k+2\mathrm{e}^\tau}(s^{(n)}-s^{(n+1)})+\frac{k}{k+2\mathrm{e}^\tau}(s^{(n)}-s^{(n+1)}+s^{(n+1)}-s^{(n+2)})\nonumber\\
	=&~\frac{\mu}{2}\left(Np^{(n)}+\frac{k}{k+2\mathrm{e}^\tau}Np^{(n+1)}-\frac{2k+2\mathrm{e}^\tau}{k+2\mathrm{e}^\tau}\right)+\frac{2k+2\mathrm{e}^\tau}{k+2\mathrm{e}^\tau}\mathcal{O}(\mu^2).
\end{align}

Then, we can calculate Eq.~(\ref{eqimicondi}) and identify the critical synergy factor over which cooperation is favored. In particular, we apply
the payoff function Eq.~(\ref{piwalk}) to Eq.~(\ref{eqimicondi}), then transform the strategy values $s^{(n)}$ into random walk values $p^{(n)}$ according to Eq.~(\ref{eqsubtrans}).
\begin{align}\label{eqimicalcu}
	&~\pi^{(0)}
	-\frac{2\mathrm{e}^\tau}{k+2\mathrm{e}^\tau}\pi^{(1)}
	-\frac{k}{k+2\mathrm{e}^\tau}\pi^{(2)}>0\nonumber
	\\
	\Leftrightarrow&\left(\frac{r}{(k+1)G}-1\right)s^{(0)}+\frac{2k}{(k+1)G}rs^{(1)}+\frac{k^2}{(k+1)G} rs^{(2)}\nonumber\\
	&-\frac{2\mathrm{e}^\tau}{k+2\mathrm{e}^\tau}
	\left(\left(\frac{r}{(k+1)G}-1\right)s^{(1)}+\frac{2k}{(k+1)G}rs^{(2)}+\frac{k^2}{(k+1)G} rs^{(3)}\right)\nonumber\\
	&-\frac{k}{k+2\mathrm{e}^\tau}
	\left(\left(\frac{r}{(k+1)G}-1\right)s^{(2)}+\frac{2k}{(k+1)G}rs^{(3)}+\frac{k^2}{(k+1)G} rs^{(4)}\right)>0\nonumber
	\\
	\Leftrightarrow&
	\left(\frac{r}{(k+1)G}-1\right)\left(s^{(0)}-\frac{2\mathrm{e}^\tau}{k+2\mathrm{e}^\tau}s^{(1)}-\frac{k}{k+2\mathrm{e}^\tau}s^{(2)}\right)\nonumber\\
	&+\frac{2k}{(k+1)G}r\left(s^{(1)}-\frac{2\mathrm{e}^\tau}{k+2\mathrm{e}^\tau}s^{(2)}-\frac{k}{k+2\mathrm{e}^\tau}s^{(3)}\right)\nonumber\\
	&+\frac{k^2}{(k+1)G} r\left(s^{(2)}-\frac{2\mathrm{e}^\tau}{k+2\mathrm{e}^\tau}s^{(3)}-\frac{k}{k+2\mathrm{e}^\tau}s^{(4)}\right)>0\nonumber
	\\
	\Leftrightarrow&
	\left(\frac{r}{(k+1)G}-1\right)\left(Np^{(0)}+\frac{k}{k+2\mathrm{e}^\tau}Np^{(1)}-\frac{2k+2\mathrm{e}^\tau}{k+2\mathrm{e}^\tau}\right)\nonumber\\
	&+\frac{2k}{(k+1)G}r\left(Np^{(1)}+\frac{k}{k+2\mathrm{e}^\tau}Np^{(2)}-\frac{2k+2\mathrm{e}^\tau}{k+2\mathrm{e}^\tau}\right)\nonumber\\
	&+\frac{k^2}{(k+1)G} r\left(Np^{(2)}+\frac{k}{k+2\mathrm{e}^\tau}Np^{(3)}-\frac{2k+2\mathrm{e}^\tau}{k+2\mathrm{e}^\tau}\right)>0.
\end{align}
The remaining part is to calculate the requested $p^{(n)}$ values. 
One stays at the starting position if not walking, so that $p^{(0)}=1$. 
Since we excluded self-loops, one cannot leave and return to the starting node within a single step, so that $p^{(1)}=0$. The $p^{(2)}$ value may vary for arbitrary networks, but is explicit on a transitive structure characterized by $k$ degree: there are $k$ choices for the first step, and the probability of each is $1/k$; for each of the first steps, the second step's probability of moving towards the starting node is $1/k$; therefore, $p^{(2)}=k\times 1/k\times 1/k=1/k$. The $p^{(3)}$ value, however, is a bit complicated and dependent on specific structures (but still conceivable). We present $p^{(3)}$ values for different spatial structures in Appendix~\ref{secappenp3}.

Applying $p^{(0)}=1$, $p^{(1)}=0$, $p^{(2)}=1/k$ but still keeping $p^{(3)}$, we can further calculate Eq.~(\ref{eqimicalcu}), 
\begin{align}\label{eqimicalcuresult}
	&~\pi^{(0)}
	-\frac{2\mathrm{e}^\tau}{k+2\mathrm{e}^\tau}\pi^{(1)}
	-\frac{k}{k+2\mathrm{e}^\tau}\pi^{(2)}>0\nonumber
	\\
	\Leftrightarrow&
	\left(\frac{r}{(k+1)G}-1\right)\left(N-\frac{2k+2\mathrm{e}^\tau}{k+2\mathrm{e}^\tau}\right)+\frac{2k}{(k+1)G}r\left(\frac{1}{k+2\mathrm{e}^\tau}N-\frac{2k+2\mathrm{e}^\tau}{k+2\mathrm{e}^\tau}\right)\nonumber\\
	&+\frac{k^2}{(k+1)G} r\left(\frac{N}{k}+\frac{k}{k+2\mathrm{e}^\tau}Np^{(3)}-\frac{2k+2\mathrm{e}^\tau}{k+2\mathrm{e}^\tau}\right)>0\nonumber
	\\
	\Leftrightarrow&~
	r>\frac{(N-2)(G-1)G^2+2(N-1)G^2\mathrm{e}^\tau}
	{N(G-1)^3 p^{(3)}+N(G-1)(G+2)-2(G-1)G^2+2G(N-G)\mathrm{e}^\tau}
	\eqqcolon r^\star,
\end{align}
which gives the analytical solution of the critical synergy factor $r^\star$, and cooperation is favored if $r>r^\star$. To simplify the parameters, we have used $k=G-1$ to replace all $k$. The critical synergy factor $r^\star$ only depends on population $N$, group size $G$, network structure $p^{(3)}$, and inertia $\tau$.

\subsection{Discussion}
\subsubsection{Connection between different updating rules by inertia}\label{secrelevance}
In the following, we analyze and discuss the potential consequences of the main result summarized by Eq.~(\ref{eqimicalcuresult}). We introduce four new notations, which help us to present the critical synergy factor compactly and elegantly. In particular, by using the following notations: $A\coloneqq (N-2)(G-1)G^2$, $B\coloneqq 2(N-1)G^2$, $C\coloneqq N(G-1)^3 p^{(3)}+N(G-1)(G+2)-2(G-1)G^2$, and $D\coloneqq 2G(N-G)$, the expression of $r^\star$ in Eq.~(\ref{eqimicalcuresult}) can be written as 
\begin{equation}
    r^\star =\frac{A+B\mathrm{e}^\tau}{C+D\mathrm{e}^\tau}.
\end{equation}

First, we note that at $\tau=0$, in the absence of inertia, we get back the original model with the imitation update, and the condition~(\ref{eqimicondi}) becomes
\begin{equation}
    \pi^{(0)}-\frac{2}{k+2}\pi^{(1)}-\frac{k}{k+2}\pi^{(2)}>0 \Leftrightarrow r>\frac{A+B}{C+D}.
\end{equation}

Second, $\tau\to-\infty$ limit leads to the condition of death-birth updating. 
According to Eq.~(\ref{eqimitation}), this limit completely ignores the importance of the focal agent, ``a randomly selected individual dies and its neighbors compete for the position.'' In the $\tau\to-\infty$ limit, the condition~(\ref{eqimicondi}) becomes
\begin{equation}
    \pi^{(0)}-\pi^{(2)}>0 \Leftrightarrow r>\frac{A}{C}.
\end{equation}

Third, in the $\tau\to+\infty$ limit, the evolution is practically blocked because the focal player's fitness is overestimated; therefore, players are reluctant to change state. In this limit, the theoretical condition~(\ref{eqimicondi}) becomes
\begin{equation}
    \pi^{(0)}-\pi^{(1)}>0 \Leftrightarrow r>\frac{B}{D}.
\end{equation}
Although we cannot say that the dynamics follows the birth-death protocol, the condition of cooperation success depicted by critical $r^\star$ is identical to the one valid for the birth-death.

\begin{figure}
	\centering
		\includegraphics[width=.9\textwidth]{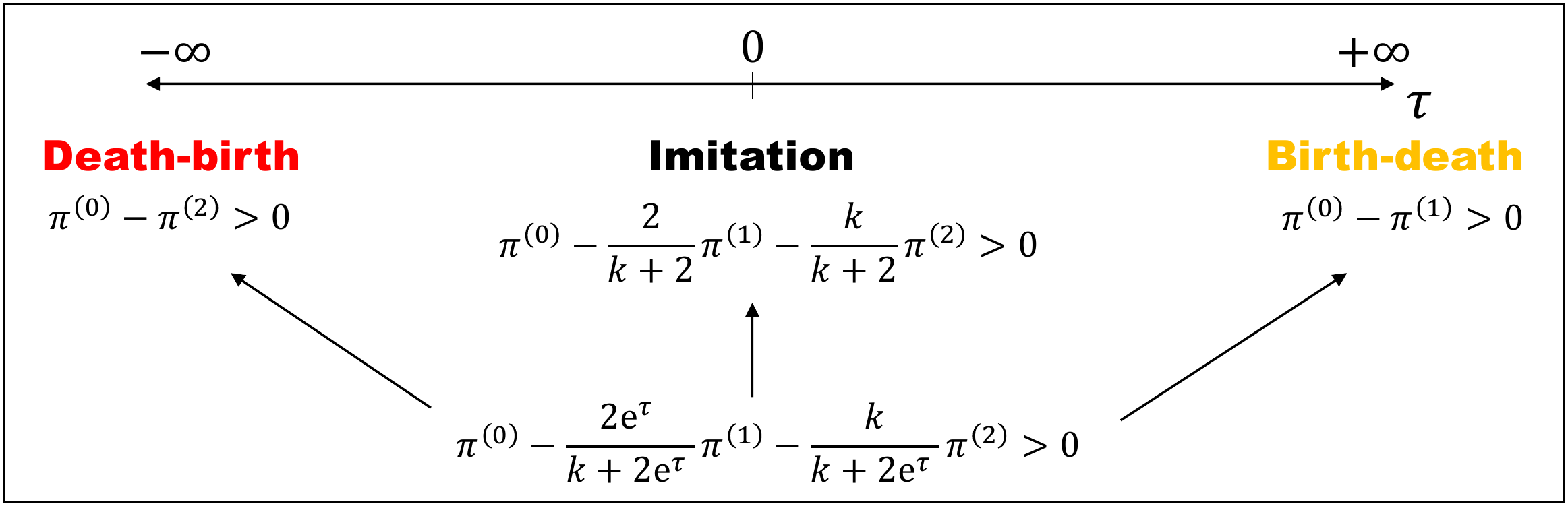}
	\caption{The diagram of the connection between inertia and the critical synergy condition for different updating rules. The cooperation success condition degenerates to the traditional imitation rule when $\tau=0$, the traditional death-birth rule when $\tau\to-\infty$, and the birth-death rule when $\tau\to+\infty$. The independent conditions for different updating rules are from Ref.~\cite{allen2014games}.} 
	\label{figbddiagram}
\end{figure}

Our findings about the possible consequence of inertia are summarized in Fig.~\ref{figbddiagram}.
For reference, the traditional condition for cooperator success under different updating rules (death-birth, birth-death, imitation) expressed by $\pi^{(n)}$ (e.g., $\pi^{(0)}-\pi^{(2)}>0$, {\it etc.}) in two-player games can be found in Ref.~\cite{allen2014games}. Notably, these expressions do not change for multiplayer games~\cite{su2019spatial}. This analysis revealed an interesting link among different strategy updating rules with the help of the inertia concept.

\subsubsection{Inertia hinders cooperation under weak selection}
We continue our discussion of the key finding by considering specific parameter values in the general form. The accurate forms of the critical synergy factor $r^\star$ under special parameters are collected in Table~\ref{tableresult}. They include the general and three special cases discussed in Section~\ref{secrelevance}. Their corresponding forms in the large population limit as $N\to+\infty$ are also presented. On the one hand, Table~\ref{tableresult} records the spatial public goods game's critical synergy factor under different inertia as well as its large population limit. On the other hand, we can say that Table~\ref{tableresult} also records the spatial public goods game's critical synergy factor under different updating rules and its large population limit.

\begin{table}[width=\linewidth,pos=h]
\caption{Main results of the critical synergy factor $r>r^\star$ for cooperation success. The general form is shown in the first line, while different cases obtained at specific values of parameters are listed in the remaining lines.}\label{tableresult}
\begin{tabular*}{\tblwidth}{@{} LLL@{} }
\toprule
\multicolumn{1}{L}{Special parameter} & 
The synergy factor $r>r^\star$ for cooperation success & 
Comment 
\\\midrule
/ &
$r^\star=\dfrac{(N-2)(G-1)G^2+2(N-1)G^2\mathrm{e}^\tau}
{N(G-1)^3 p^{(3)}+N(G-1)(G+2)-2(G-1)G^2+2G(N-G)\mathrm{e}^\tau}$ &
General
\\
$\tau\to -\infty$ & 
$r^\star=
\dfrac{(N-2)G^2}
{N(G-1)^2 p^{(3)}+N(G+2)-2G^2}$ & 
Death-birth (DB)
\rule{0em}{2em}\\
$\tau=0$ &
$r^\star=\dfrac{(N-2)(G-1)G^2+2(N-1)G^2}
{N(G-1)^3 p^{(3)}+N(G-1)(G+2)-2(G-1)G^2+2G(N-G)}$ &
Imitation (IM)
\rule{0em}{2em}\\
$\tau\to +\infty$ & 
$r^\star=
\dfrac{(N-1)G}
{N-G}$ & 
Birth-death (BD)
\rule{0em}{2em}\\
$N\to +\infty$ & 
$r^\star=\dfrac{(G-1)G^2+2G^2\mathrm{e}^\tau}
{(G-1)^3 p^{(3)}+(G-1)(G+2)+2G\mathrm{e}^\tau}$ & 
General in large population
\rule{0em}{2em}\\
$N\to +\infty$, $\tau\to -\infty$ & 
$r^\star=\dfrac{G^2}
{(G-1)^2 p^{(3)}+G+2}$ & 
DB in large population
\rule{0em}{2em}\\
$N\to +\infty$, $\tau=0$ & 
$r^\star=\dfrac{(G-1)G^2+2G^2}
{(G-1)^3 p^{(3)}+(G-1)(G+2)+2G}$ & 
IM in large population
\rule{0em}{2em}\\
$N\to +\infty$, $\tau\to +\infty$ & 
$r^\star=
G$ & 
BD in large population
\rule{0em}{2em}\\
\bottomrule
\end{tabular*}
\end{table}

We already stressed that the presented results are generally valid because the specific features of interaction topology are encapsulated in the values of $G$ and $p^{(3)}$. In the following, we consider different neighborhoods for public goods games played on the square lattice, while other lattices are discussed in Appendix~B. The calculation for $p^{(3)}$ values are given in Appendix~\ref{secappenp3}.
Table~\ref{tablevalue} summarizes the specific results of Table~\ref{tableresult} obtained for different group sizes, where we present the most popular settings, including $G=5$ (von~Neumann neighborhood), $G=9$ (Moore neighborhood), $G=13$, and $G=25$. 

\begin{table}[width=\linewidth,pos=h]
\caption{Specified results of the critical synergy factor $r>r^\star$ for cooperation success on the square lattice using different group sizes. The cases of $G=5$ (von~Neumann neighborhood), $G=9$ (Moore neighborhood), $G=13$, and $G=25$ are presented. See Appendix~\ref{secappenp3} for the $p^{(3)}$ value in each case.}\label{tablevalue}
\begin{tabular*}{\tblwidth}{@{} LLL@{} }
\toprule
\multicolumn{1}{L}{Special parameter} & 
$G=5$, $p^{(3)}=0$ & 
$G=9$, $p^{(3)}=3/64$
\\\midrule
/ &
$r^\star=\dfrac{50N-100+25(N-1)\mathrm{e}^\tau}
{14N-100+5(N-5)\mathrm{e}^\tau}$ &
$r^\star=\dfrac{324N-648+81(N-1)\mathrm{e}^\tau}
{56N-648+9(N-9)\mathrm{e}^\tau}$
\\
$\tau\to -\infty$ & 
$r^\star=
\dfrac{25N-50}
{7N-50}$ & 
$r^\star=\dfrac{81N-162}
{14N-162}$
\rule{0em}{2em}\\
$\tau=0$ &
$r^\star=
\dfrac{75N-125}{19N-125}$ &
$r^\star=
\dfrac{405N-729}{65N-729}$
\rule{0em}{2em}\\
$\tau\to +\infty$ & 
$r^\star=
\dfrac{5(N-1)}
{N-5}$ & 
$r^\star=
\dfrac{9(N-1)}
{N-9}$
\rule{0em}{2em}\\
$N\to +\infty$ & 
$r^\star=\dfrac{50+25\mathrm{e}^\tau}
{14+5\mathrm{e}^\tau}$ & 
$r^\star=\dfrac{324+81\mathrm{e}^\tau}
{56+9\mathrm{e}^\tau}$
\rule{0em}{2em}\\
$N\to +\infty$, $\tau\to -\infty$ & 
$r^\star=\dfrac{25}
{7}\approx 3.5714$ & 
$r^\star=\dfrac{81}
{14}\approx 5.7857$
\rule{0em}{2em}\\
$N\to +\infty$, $\tau=0$ & 
$r^\star=
\dfrac{75}{19}\approx 3.9474$ & 
$r^\star=
\dfrac{81}{13}\approx 6.2308$
\rule{0em}{2em}\\
$N\to +\infty$, $\tau\to +\infty$ & 
$r^\star=
5$ & 
$r^\star=
9$
\rule{0em}{2em}\\
\bottomrule

 & 
$G=13$, $p^{(3)}=5/144$ & 
$G=25$, $p^{(3)}=1/48$ 
\rule{0em}{1.1em}\\\midrule
/ &
$r^\star=\dfrac{1014N-2028+169(N-1)\mathrm{e}^\tau}
{120N-2028+13(N-13)\mathrm{e}^\tau}$ &
$r^\star=\dfrac{7500N-15000+625(N-1)\mathrm{e}^\tau}
{468N-15000+25(N-25)\mathrm{e}^\tau}$
\\
$\tau\to -\infty$ & 
$r^\star=
\dfrac{169N-338}
{20N-338}$ & 
$r^\star=\dfrac{625N-1250}
{39N-1250}$
\rule{0em}{2em}\\
$\tau=0$ &
$r^\star=
\dfrac{1183N-2197}{133N-2197}$ &
$r^\star=
\dfrac{8125N-15625}{493N-15625}$
\rule{0em}{2em}\\
$\tau\to +\infty$ & 
$r^\star=
\dfrac{13(N-1)}
{N-13}$ & 
$r^\star=
\dfrac{25(N-1)}
{N-25}$
\rule{0em}{2em}\\
$N\to +\infty$ & 
$r^\star=\dfrac{1014+169\mathrm{e}^\tau}
{120+13\mathrm{e}^\tau}$ & 
$r^\star=\dfrac{7500+625\mathrm{e}^\tau}
{468+25\mathrm{e}^\tau}$
\rule{0em}{2em}\\
$N\to +\infty$, $\tau\to -\infty$ & 
$r^\star=\dfrac{507}
{60}=8.45$ & 
$r^\star=\dfrac{625}
{39}\approx 16.0256$
\rule{0em}{2em}\\
$N\to +\infty$, $\tau=0$ & 
$r^\star=
\dfrac{169}{19}\approx 8.8947$ & 
$r^\star=
\dfrac{8125}{493}\approx 16.4807$
\rule{0em}{2em}\\
$N\to +\infty$, $\tau\to +\infty$ & 
$r^\star=
13$ & 
$r^\star=
25$
\rule{0em}{2em}\\
\bottomrule
\end{tabular*}
\end{table}

\begin{figure}
	\centering
		\includegraphics[width=.9\textwidth]{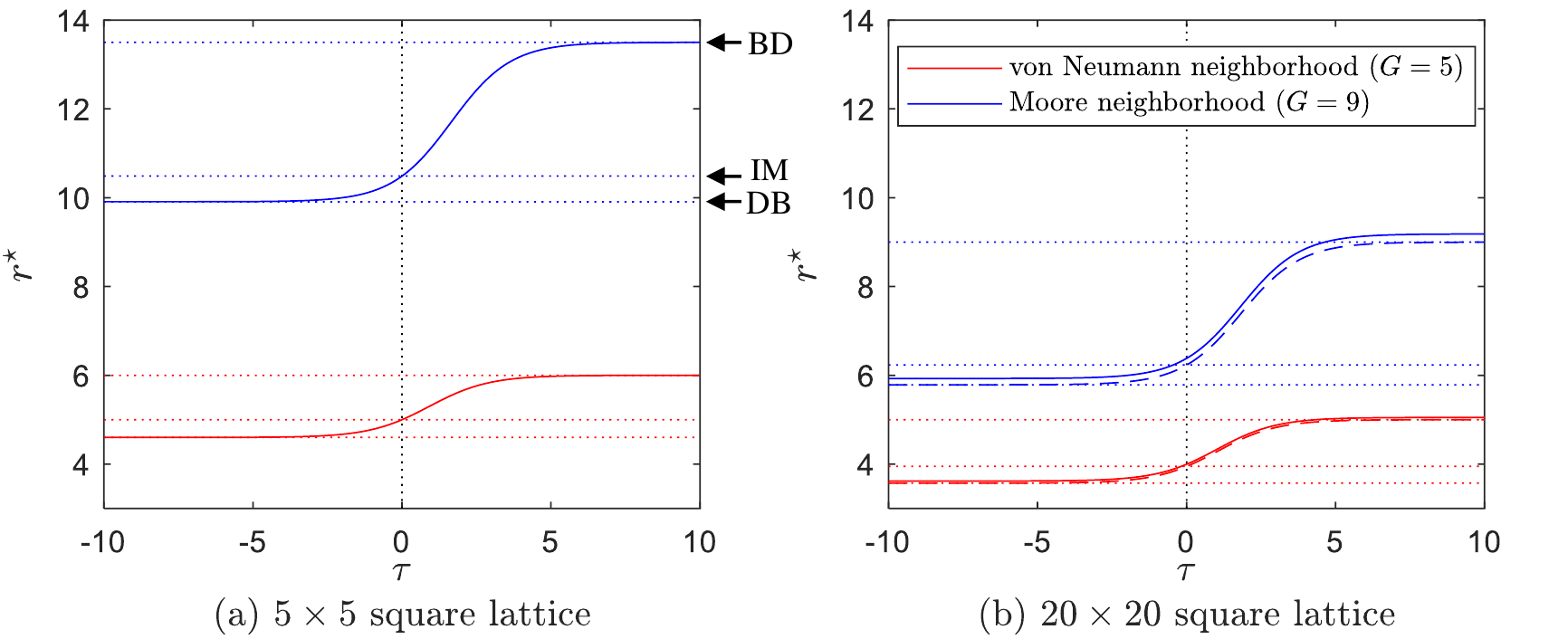}
	\caption{The analytical solution for the critical synergy factor $r^\star$ as a function of inertia $\tau$. (a) A $5\times 5$ square lattice, where $N=25$. The two solid lines represent the functions under $G=5$ and $G=9$, as shown in the legend. The three horizontal dotted lines around each function from bottom to top are the $r^\star$ values under death-birth (DB), imitation (IM), and birth-death (BD) updating rules. (b) A $20\times 20$ square lattice, $N=400$. The two solid lines represent the functions under $G=5$ and $G=9$. In addition, the dashed lines near them are the functions under a large population limit $N\to+\infty$. The three horizontal dotted lines around each function from bottom to top are the $N\to+\infty$ large population's $r^\star$ values under DB, IM, and BD updating rules.}
	\label{figtheor}
\end{figure}

Next, we illustrate how the critical synergy factor depends on $\tau$ by calculating the formulas collected in Table~\ref{tableresult}. Our results are shown in Fig.~\ref{figtheor}, where we present $r^\star$ values for $G=5$ and $G=9$ group sizes. The two panels represent small and significantly large system sizes. Besides the analytical values, we also present three horizontal lines marking the threshold levels calculated for death-birth, imitation, and birth-death updating rules. These values fit well with the corresponding function curves at $\tau=-10$, $0$, $10$, but the limit $r^\star$ values have already reached around $|\tau| \approx 5$ inertia level. The right panel shows the same quantities for $N=400$ system size. We also plot the accurate $r^\star$ values obtained for the $N\to+\infty$ limit. These curves are in the vicinity of $N=400$ curves, indicating that this system size can be considered large.
In both Fig.~\ref{figtheor}(a) and Fig.~\ref{figtheor}(b), as inertia $\tau$ increases, the critical synergy factor $r^\star$ increases, which establishes a more demanding $r>r^\star$ condition; that is, an increase in $\tau$ disfavors cooperation. This observation is generally valid because $\partial r^\star/\partial \tau>0$ in our analytical formula of Eq.~(\ref{eqimicalcuresult}). Therefore, the presence of inertia always hinders cooperation in spatial public goods games under weak selection strength. 

\section{Numerical results}\label{secnume}
To confirm the analytical results, we provide numerical calculations. We fix $\kappa=0.1$, and $c=0.001$ or $c=0.01$, such that $\delta=c/\kappa=0.01$ or $\delta=c/\kappa=0.1$, which can be considered as a weak selection~\cite{allen2017evolutionary,ohtsuki2006simple}. Importantly, this parametrization differs from previous numerical works, which assumed $c=1$ such that their selection strength is intermediate.

\begin{figure}
	\centering
		\includegraphics[width=\textwidth]{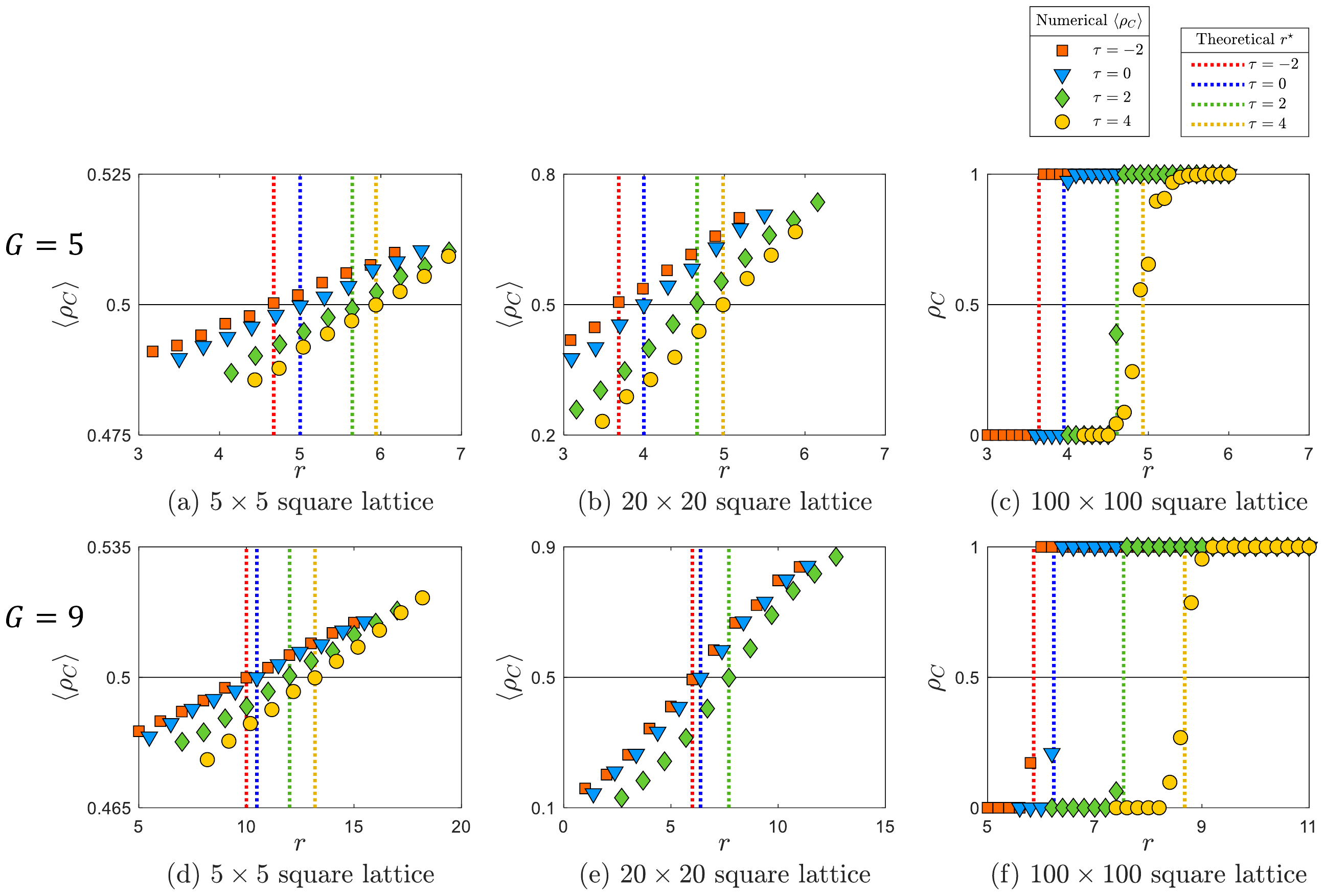}
	\caption{Numerical simulations confirm our theoretical predictions. Symbols show the average cooperation frequency $\langle\rho_C\rangle$ of Monte Carlo simulations in dependence of $r$ for different $\tau$ values, as indicated in the legend. Vertical dashed lines represent theoretical critical synergy factor $r^\star$ over which $\langle\rho_C\rangle>0.5$. In all panels, $\kappa=0.1$. Panel~(a) and (d): $N=25$, $c=0.001$, and $\langle\rho_C\rangle$ is the average over 2000000 runs. Panel~(b) and (e): $N=400$, $c=0.001$, and $\langle\rho_C\rangle$ is the average over 10000 runs. Panel~(c) and (f): $N=10000$, $c=0.01$, and $\rho_C$ is the result of a single run. The top row shows the results for $G=5$, and the bottom row shows $G=9$ group size.}
	\label{fignume}
\end{figure}

The cooperation level in the system is denoted by $\rho_C=N_C/N$, where $N_C$ is the number of cooperative agents in the system. Initially, we randomly assign the strategy of cooperation or defection to each agent, such that the starting cooperation level is $\rho_C\approx 0.5$. We let agents play games and update their strategies according to Section~\ref{secmodel} and end a run when $N_C=N$ or $N_C=0$. Then, $\rho_C=1$ or $\rho_C=0$ is recorded as a result of a single run. If the system does not fixate before the max time step, we record the $0<\rho_C<1$ cooperation level obtained in the final max step. For $5\times 5$ and $20\times 20$ lattices, we allow up to $t=400000$ full MC steps~\cite{allen2017evolutionary}, and for the $100\times 100$ lattice, we allow up to $t=10000$ full MC steps. We repeat the abovementioned simulation multiple times (see the caption of Fig.~\ref{fignume} for details) and average their results, denoted by $\langle\rho_C\rangle$.

The cooperation success condition that we deduced in Section~\ref{sectheo} means that the expected cooperation level $\langle\rho_C\rangle>N_C/N$ when starting with $N_C$ cooperators. For simplicity, we started with a single cooperator when deducing the theoretical condition. However, in numerical simulations, we start with $N_C/N\approx 0.5$ by randomly assigning strategies; therefore, cooperation success means $\langle\rho_C\rangle>1/2$.

Figure~\ref{fignume} shows the average cooperation level $\langle\rho_C\rangle$ as a function of synergy factor $r$ on square lattices with different sizes ($5\times 5$, $20\times 20$, and $100\times 100$ for the three columns) and neighbors (von~Neumann and Moore neighborhood for the two rows). Naturally, as $r$ increases, the cooperation level $\langle\rho_C\rangle$ increases. Substituting $G=5$, $N=25$, $400$, $10000$, $\tau=-2$, $0$, $2$, $4$, and $p^{(3)}=0$, $3/64$ (depending on different panels of Fig.~\ref{fignume}) into Eq.~(\ref{eqimicalcuresult}), we can calculate the critical synergy factor $r^\star$. For different inertia values $\tau=-2$, $0$, $2$, $4$, we have $r^\star\approx 4.6719$, $5$, $5.6461$, $5.9387$ in Fig~\ref{fignume}(a), $r^\star\approx 3.6846$, $3.9967$, $4.6585$, $4.9811$ in Fig~\ref{fignume}(b), $r^\star\approx 3.6392$, $3.9493$, $4.6094$, $4.9323$ in Fig~\ref{fignume}(c), $r^\star\approx 10.0003$, $10.4866$, $12.0132$, $13.1866$ in Fig~\ref{fignume}(d), $r^\star\approx 5.9980$, $6.3817$, $7.7011$, $8.8530$ in Fig~\ref{fignume}(e) (due to very slow relaxation, the case of $\tau=4$ is not shown, but the remaining cases are enough to draw the conclusion), and $r^\star\approx 5.8597$, $6.2366$, $7.5373$, $8.6783$ in Fig~\ref{fignume}(f). When $r>r^\star$, the simulation predicts cooperation level $\langle\rho_C\rangle>0.5$, which validates the theoretical condition of cooperation success. 

\begin{figure}
	\centering
		\includegraphics[width=.6\textwidth]{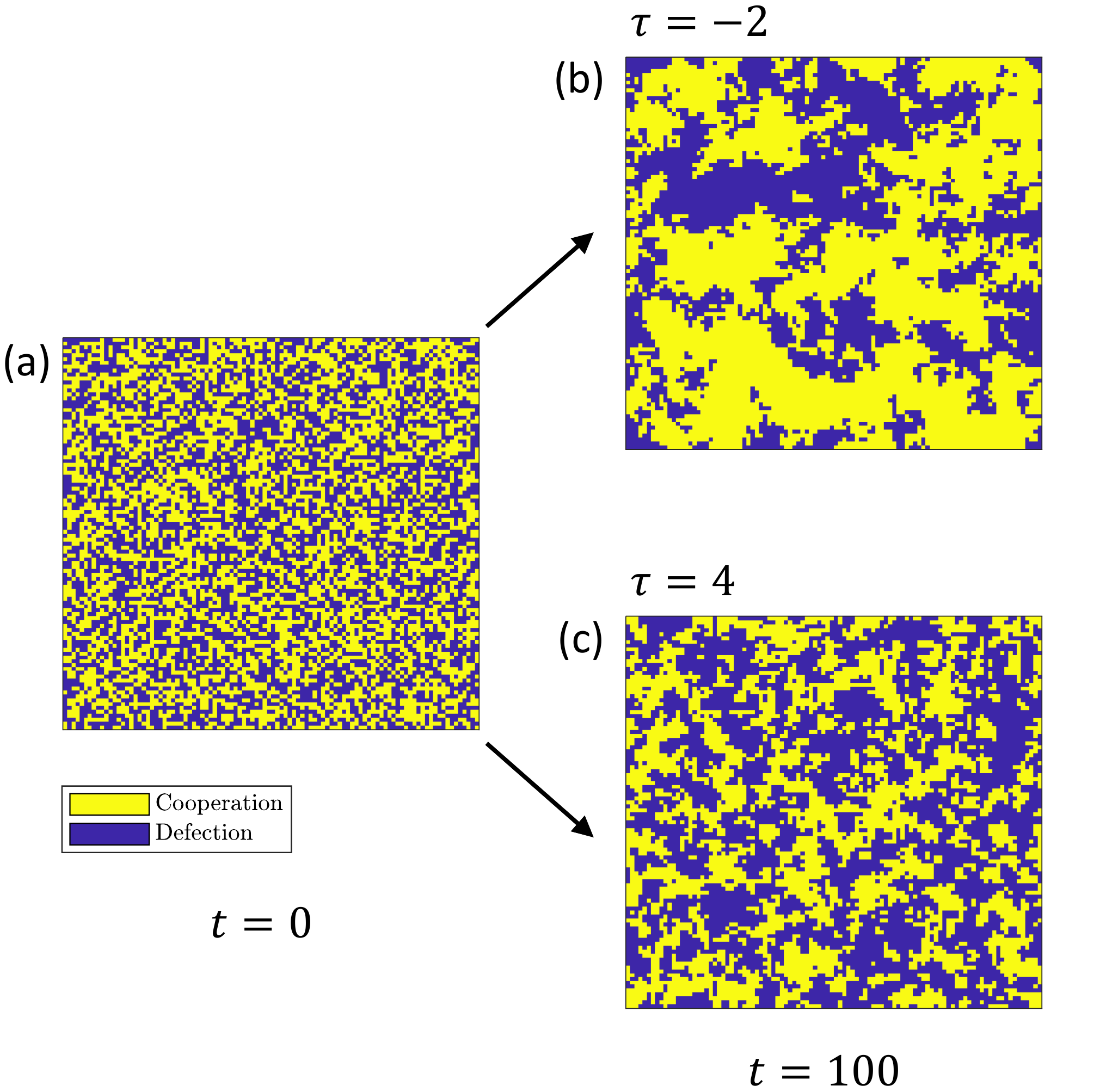}
	\caption{Time evolution of strategy distribution under weak selection in the presence of inertia, where $G=5$, $r=4$. Panel~(a) shows the random initial state of the $100\times 100$ system. Panel~(b) and (c) show the evolving patterns after 100 MC steps for different $\tau$ values, as indicated. When a large $\tau$ value is applied, the voter-model-like random coarsening remains slow, which offers many interfaces, hence many opportunities for defectors to exploit the rival strategy. In the alternative case, shown in panel~(b), the local coarsening process is faster, which allows cooperators to condense, hence protecting themselves from defectors via spatial reciprocity.}
	\label{figshot}
\end{figure}

To better understand why inertia undermines the evolution of cooperation under weak selection, we illustrate how the spatial distribution of strategies evolves for different $\tau$ values. Figure~\ref{figshot} shows this process on a $100\times 100$ lattice. At $t=0$, the two strategies are randomly distributed, as shown in Fig.~\ref{figshot}(a). We note that without considering payoff values, a logarithmically slow coarsening would start, reminding us of the voter model dynamics~\cite{dornic_prl01}. Evidently, the weak selection in the evolutionary process alters this dynamics gently and results in different evolutionary paths for different $\tau$ values. Panel~(b) and panel~(c) depict the stages after $t=100$ steps for $\tau = -2$ and $\tau=4$, respectively. It is clear that the coarsening process remains very slow for $\tau=4$; hence competing strategies can only form small islands separated by many interfaces. This quasi-random state supports defectors who can easily find cooperator neighbors to exploit. In the opposite case, shown in panel~(b), the local ordering is significantly faster; hence cooperators can form larger domains. This ordering makes it possible for network reciprocity to work, which provides a better condition for the evolution of cooperation.

\section{Conclusion}\label{secconclu}
While some earlier works already studied the concept of inertia~\cite{du2012effects,chang2018cooperation,liu_rr_epl10}, its impact on the evolutionary process in spatial populations under weak selection remained largely unexplored. Motivated by this, we revisited this idea by aiming for analytical results about how it affects the evolutionary process.
With the help of the IBD method, we deduced the accurate form of the critical synergy factor $r^\star$ in spatial public goods games with inertia, where cooperation succeeds if $r>r^\star$. We find that $r^\star$ only depends on population $N$, group size $G$, and spatial structure $p^{(3)}$. The results hold under the weak selection strength limit, which is equivalent to the contribution being much smaller than the noise parameter, $0<c\ll \kappa$.

As an interesting observation, we find that inertia $\tau$ links the death-birth ($\tau\to-\infty$), imitation ($\tau=0$), and birth-death ($\tau\to+\infty$) updating rules in the condition of cooperation success. More precisely, the critical synergy condition obtained for imitation can reproduce the conditions for death-birth and birth-death rules if we formally take the $\tau \to -\infty$ and $\tau \to +\infty$ limit of the inertia parameter. We generally find that the presence of inertia is detrimental and makes the evolution of cooperation harder. Technically, the $r^\star$ critical synergy factor is a growing function of $\tau$. This observation is robust and remains valid independently of the group sizes used in the social dilemma.

Monte Carlo simulations confirm that the average cooperation level (fraction of cooperators) exceeds the initial value 0.5 when $r>r^\star$. As inertia increases, cooperation is disfavored, and the critical synergy factor $r^\star$ is elevated. When monitoring the time evolution of the strategy distributions, we can observe that smaller inertia allows cooperators to form larger clusters and support each other, thus favoring cooperation. Meanwhile, larger inertia slows the evolution, making dynamics close to the neutral drift. As a result, the spatial distribution remains quasi-random, which provides a supportive environment for defectors.

\section*{Data availability}
No data was used for the research described in the article.

\section*{Acknowledgments}
A.S. was supported by the National Research, Development and Innovation Office (NKFIH) under Grant No.~K142948.

\appendix
\renewcommand\thefigure{\Alph{section}\arabic{figure}} 
\section{Pairwise comparison}\label{secappenfermi}
\setcounter{figure}{0}
An alternative strategy updating protocol is the pairwise comparison. In this case, a focal player $i$ and one of its neighbors $j\in\Omega_i\backslash\{i\}$ are randomly selected during an elementary step. Agent $i$ imitates the strategy of $j$ with a probability depending on their payoff difference. The most popular choice for this probability function is the so-called Fermi function~\cite{szabo1998evolutionary}; agent $i$ learns $j$'s strategy $s_j$ with the probability
\begin{equation}\label{eqfermi}
	W_\text{Fermi}(s_i\gets s_j)
	=\frac{F_j}{F_i+F_j}
	=\frac{\exp(\pi_j/\kappa)}{\exp(\tau+\pi_i/\kappa)+\exp(\pi_j/\kappa)}
	=\frac{1}{1+\exp(\tau+(\pi_i-\pi_j)/\kappa)};
\end{equation}
otherwise, agent $i$ keeps its original strategy $s_i$. The inertia parameter has been introduced in Eq.~(\ref{eqfermi}), which degenerates to the classic Fermi function when $\tau=0$.

For the condition of cooperation success, we need to write $\mathcal{B}_1$ and $\mathcal{D}_1$ for a single initial cooperator, similar to the previously discussed updating rules.
\begin{subequations}\label{eqbdfermi}
	\begin{align}
		\mathcal{B}_1&=\sum_{i\in \Omega_1\backslash\{1\}}\frac{1}{N}W_\text{Fermi}(s_i\gets s_1)
		=\sum_{i\in \Omega_1\backslash\{1\}}\frac{1}{N}
		\frac{1}{1+\exp(\tau+(\pi_i-\pi_1)/\kappa)},
		\\
		\mathcal{D}_1&=\frac{1}{N}\sum_{j\in \Omega_1\backslash\{1\}}W_\text{Fermi}(s_1\gets s_j)
		=\frac{1}{N}\sum_{j\in \Omega_1\backslash\{1\}}\frac{1}{1+\exp(\tau+(\pi_1-\pi_j)/\kappa)}.
	\end{align}
\end{subequations}
Then, calculating the condition for cooperation success leads to
\begin{align}\label{condigroupfermi}
    &\left\langle\frac{\partial}{\partial\delta}(\mathcal{B}_1-\mathcal{D}_1)\right\rangle_{\begin{smallmatrix}\delta=0\\s_1=1\end{smallmatrix}}>0\nonumber\\
    \Leftrightarrow&~
    \frac{1}{N}\left\langle-\sum_{i\in \Omega_1\backslash\{1\}}(\pi_i-\pi_1)\right\rangle_{\begin{smallmatrix}\delta=0\\s_1=1\end{smallmatrix}}
    -\frac{1}{N}\left\langle-\sum_{j\in \Omega_1\backslash\{1\}}(\pi_1-\pi_j)\right\rangle_{\begin{smallmatrix}\delta=0\\s_1=1\end{smallmatrix}}
    >0\nonumber\\
    \Leftrightarrow&~
    \pi^{(0)}-\pi^{(1)}>0,
\end{align}
which is independent of $\tau$. 

Again, considering the original substitution~(\ref{eqsub}) and performing the calculation,
\begin{align}\label{eqfermiresult}
    &~\pi^{(0)}-\pi^{(1)}>0\nonumber\\
    \Leftrightarrow&
    \left(\frac{r}{(k+1)G}-1\right)s^{(0)}+\frac{2k}{(k+1)G}rs^{(1)}+\frac{k^2}{(k+1)G} rs^{(2)}\nonumber\\
    &-\left(\left(\frac{r}{(k+1)G}-1\right)s^{(1)}+\frac{2k}{(k+1)G}rs^{(2)}+\frac{k^2}{(k+1)G} rs^{(3)}\right)>0\nonumber\\
    \Leftrightarrow&
    \left(\frac{r}{(k+1)G}-1\right)\left(s^{(0)}-s^{(1)}\right)
    +\frac{2k}{(k+1)G}r\left(s^{(1)}-s^{(2)}\right)
    +\frac{k^2}{(k+1)G}r\left(s^{(2)}-s^{(3)}\right)>0\nonumber\\
    \Leftrightarrow&
    \left(\frac{r}{(k+1)G}-1\right)\left(Np^{(0)}-1\right)
    +\frac{2k}{(k+1)G}r\left(Np^{(1)}-1\right)
    +\frac{k^2}{(k+1)G}r\left(Np^{(2)}-1\right)>0\nonumber\\
    \Leftrightarrow&
    \left(\frac{r}{(k+1)G}-1\right)\left(N-1\right)
    -\frac{2k}{(k+1)G}r
    +\frac{k^2}{(k+1)G}r\left(\frac{N}{k}-1\right)>0\nonumber\\
    \Leftrightarrow&~
    r>\frac{(N-1)G}{N-G},
\end{align}
which is the cooperation success condition in the pairwise comparison rule and is independent of $\tau$. In particular, for a large population, $N\to+\infty$, Eq.~(\ref{condigroupfermi}) becomes
\begin{equation}\label{eqfermiresultlarge}
    r>G.
\end{equation}
We stress that the specific feature of interaction topology has only a role in $G$. Therefore, the independence of $\tau$ is generally valid for all transitive graphs, hence for all lattices.

\begin{figure}
	\centering
		\includegraphics[width=.8\textwidth]{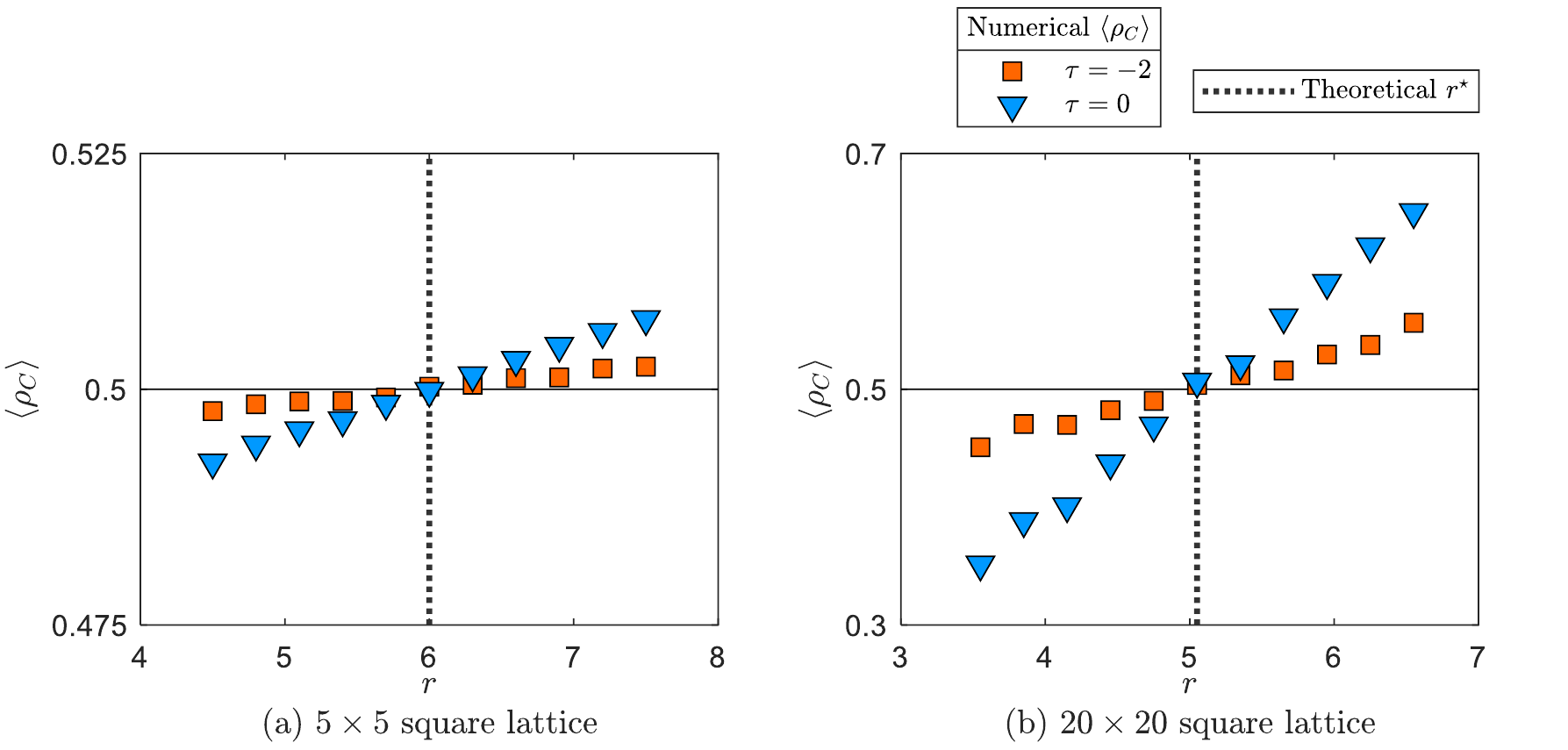}
	\caption{Numerical simulations and the prediction of theoretical analysis under pairwise comparison rule. In both panels, $c=0.001$, $\kappa=0.1$, $G=5$. (a) $N=25$ and 2000000 runs, (b) $N=400$ and 10000 runs. These results suggest that inertia has no impact on the critical threshold of cooperation success.}
	\label{figfermi}
\end{figure}

We show Monte Carlo simulations for pairwise comparison in Fig.~\ref{figfermi}. According to both theoretical and numerical observations, while the presence of inertia hinders cooperation in imitation learning and relates the conditions obtained for death-birth and birth-death protocols, it has no impact on the conditions for the pairwise comparison rule. This is valid in the weak selection limit.

\section{Triangular and hexagonal lattices}\label{secappenother}
The original IBD method can be applied to any transitive structure. As an example, we extend our results to triangular and hexagonal lattices. 

We only show here the cases of the smallest groups, but the calculations can be extended in a similar way as we did on square lattices. Accordingly, on a triangular lattice, a player has $k=3$ neighbors, hence groups of $G=4$ for the public goods game, while on a hexagonal lattice $k=6$, hence $G=7$. The calculation of $p^{(3)}$ values are explained in Appendix~\ref{secappenp3}. Our main findings are listed in Table~\ref{tablevalue2}. These $r^\star$ values confirm our general conclusion. Namely, the evolution of cooperation becomes more demanding for large inertia, while the easiest condition can be reached when this parameter is negative.

\begin{table}[width=\linewidth,pos=h]
\caption{Supplement results of the synergy factor $r>r^\star$ for cooperation success. The cases of $G=4$ (triangular lattices) and $G=7$ (hexagonal lattices) are presented. See Appendix~\ref{secappenp3} for the $p^{(3)}$ value in each case.}\label{tablevalue2}
\begin{tabular*}{\tblwidth}{@{} LLL@{} }
\toprule
\multicolumn{1}{L}{Special parameter} & 
$G=4$, $p^{(3)}=0$ & 
$G=7$, $p^{(3)}=1/18$
\\\midrule
/ &
$r^\star=\dfrac{24N-48+16(N-1)\mathrm{e}^\tau}
{9N-48+4(N-4)\mathrm{e}^\tau}$ &
$r^\star=\dfrac{147N-294+49(N-1)\mathrm{e}^\tau}
{33N-294+7(N-7)\mathrm{e}^\tau}$
\\
$\tau\to -\infty$ & 
$r^\star=
\dfrac{8N-16}
{3N-16}$ & 
$r^\star=\dfrac{49N-98}
{11N-98}$
\rule{0em}{2em}\\
$\tau=0$ &
$r^\star=
\dfrac{40N-64}{13N-64}$ &
$r^\star=
\dfrac{196N-343}{40N-343}$
\rule{0em}{2em}\\
$\tau\to +\infty$ & 
$r^\star=
\dfrac{4(N-1)}
{N-4}$ & 
$r^\star=
\dfrac{7(N-1)}
{N-7}$
\rule{0em}{2em}\\
$N\to +\infty$ & 
$r^\star=\dfrac{24+16\mathrm{e}^\tau}
{9+4\mathrm{e}^\tau}$ & 
$r^\star=\dfrac{147+49\mathrm{e}^\tau}
{33+7\mathrm{e}^\tau}$
\rule{0em}{2em}\\
$N\to +\infty$, $\tau\to -\infty$ & 
$r^\star=\dfrac{8}
{3}\approx 2.6667$ & 
$r^\star=\dfrac{49}
{11}\approx 4.4545$
\rule{0em}{2em}\\
$N\to +\infty$, $\tau=0$ & 
$r^\star=
\dfrac{40}{13}\approx 3.0769$ & 
$r^\star=
\dfrac{49}{10}= 4.9$
\rule{0em}{2em}\\
$N\to +\infty$, $\tau\to +\infty$ & 
$r^\star=
4$ & 
$r^\star=
7$
\rule{0em}{2em}\\
\bottomrule

\end{tabular*}

\end{table}

\section{The calculation of $p^{(3)}$ values}\label{secappenp3}
\setcounter{figure}{0}
As mentioned in the main text, $p^{(3)}$ denotes the probability that after three steps of random walking, we arrive back at the starting node on the lattice. The calculation of $p^{(3)}$ values, $p^{(3)}=\sum_{j\in\Omega_i\backslash\{i\}}\sum_{ \ell\in\Omega_j\backslash\{j\}}\sum_{i\in\Omega_\ell\backslash\{\ell\}}/k^3$, is straightforward and can be done in a similar but less intuitive way how $p^{(2)}$ is derived.
To make the calculation intuitive, previous work introduced new parameters, like the assortment coefficient~\cite{su2019spatial}, but the approach did not reduce the number of independent parameters. Alternatively, here we provide the calculation of $p^{(3)}$ by direct visualization.

\begin{figure}
	\centering
		\includegraphics[width=\textwidth]{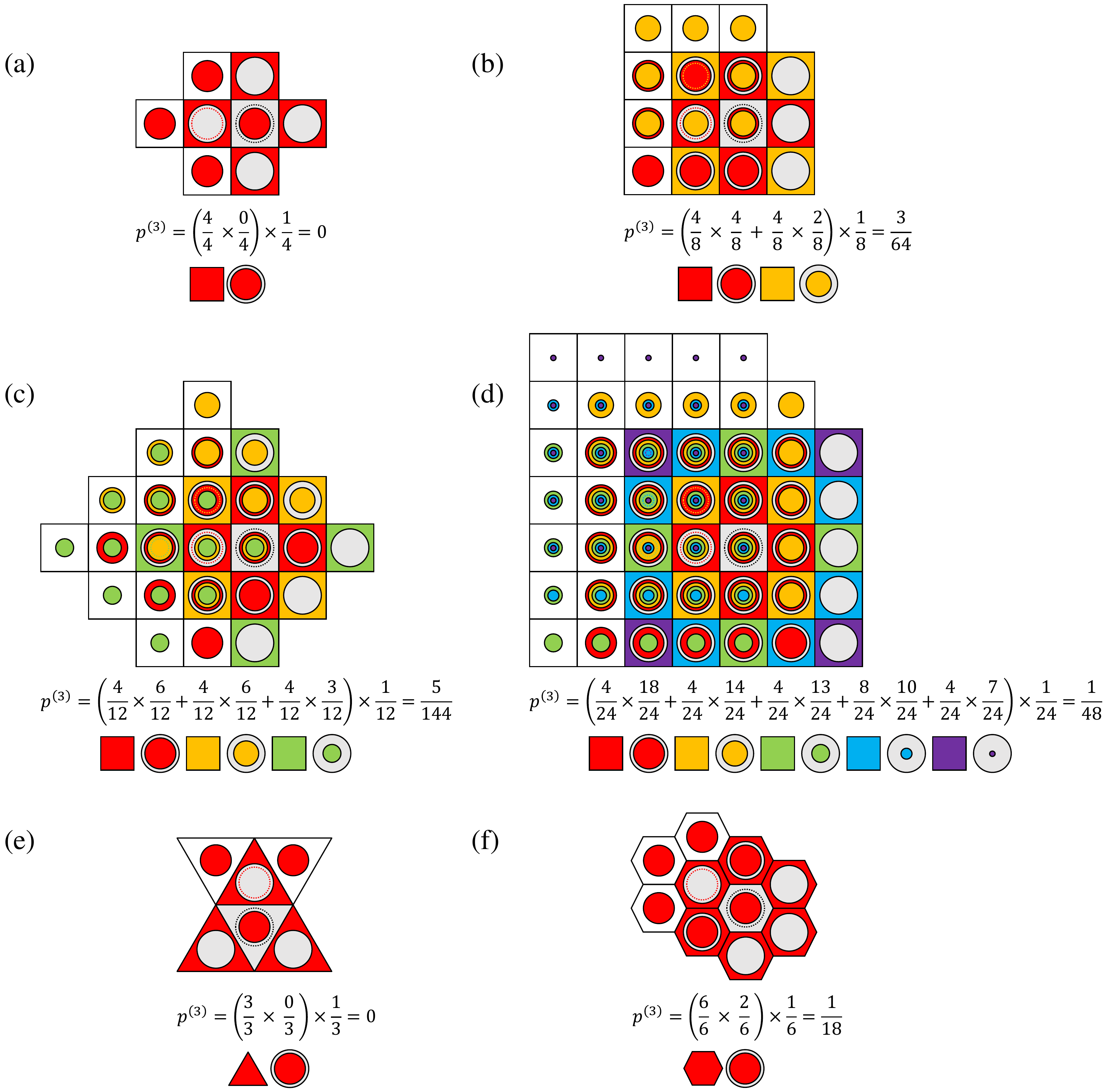}
	\caption{Graphical illustration of how $p^{(3)}$ is calculated. Panels~(a) to (d) show square lattice host graphs by using $G=5$, $G=9$, $G=13$, and $G=25$ group sizes, respectively. Panel~(e) illustrates a triangle lattice with $G=4$ group size, while panel~(f) shows a hexagonal lattice with $G=7$ group size. The general process is the following: we identify the positions of the first-order neighbors around the starting node and discuss them separately. For each position, we check how many second-order neighbors are still the starting node's neighbor. Finally, the return to the starting node happens with probability $1/k$.}
	\label{figp3}
\end{figure}

Figure~\ref{figp3} illustrates the calculation of $p^{(3)}$ on different lattices. In a word, it is equivalent to counting whether $i$'s neighbor's neighbor is still $i$'s neighbor. For the first step, different neighbors may lead to different possibilities for the second steps. We take panel~(b), a square lattice with Moore neighborhood, as an example. If the first step ends at the north, south, east, or west neighbor (with probability $4/8$), then there are four available second steps towards the starting node's neighbors; in this case (with probability $4/8$), the probability that the third step ends at the starting node is $1/8$. If the first step ends at the north-east, north-west, south-east, and south-west neighbor (with probability $4/8$), there are two available second steps toward the starting node's neighbors; in this way (with probability $2/8$), the third step ends at the starting node with probability $1/8$. To sum up, $p^{(3)}=(4/8\times 4/8+4/8\times 2/8)\times 1/8=3/64$. Similarly, we can calculate other group sizes, or other lattices, as shown in Fig.~\ref{figp3}. 

\bibliographystyle{model1-num-names}

\begin{thebibliography}{55}
\expandafter\ifx\csname natexlab\endcsname\relax\def\natexlab#1{#1}\fi
\providecommand{\url}[1]{\texttt{#1}}
\providecommand{\href}[2]{#2}
\providecommand{\path}[1]{#1}
\providecommand{\DOIprefix}{doi:}
\providecommand{\ArXivprefix}{arXiv:}
\providecommand{\URLprefix}{URL: }
\providecommand{\Pubmedprefix}{pmid:}
\providecommand{\doi}[1]{\href{http://dx.doi.org/#1}{\path{#1}}}
\providecommand{\Pubmed}[1]{\href{pmid:#1}{\path{#1}}}
\providecommand{\bibinfo}[2]{#2}
\ifx\xfnm\relax \def\xfnm[#1]{\unskip,\space#1}\fi
\bibitem[{Maynard~Smith(1982)}]{maynard_82}
\bibinfo{author}{J.~Maynard~Smith}, \bibinfo{title}{Evolution and the Theory of
  Games}, \bibinfo{publisher}{Cambridge University Press},
  \bibinfo{address}{Cambridge, U.K.}, \bibinfo{year}{1982}.
\bibitem[{Sigmund(2010)}]{sigmund_10}
\bibinfo{author}{K.~Sigmund}, \bibinfo{title}{The Calculus of Selfishness},
  \bibinfo{publisher}{Princeton University Press}, \bibinfo{address}{Princeton,
  NJ}, \bibinfo{year}{2010}.
\bibitem[{Szab{\'o} and F{\'a}th(2007)}]{szabo_pr07}
\bibinfo{author}{G.~Szab{\'o}}, \bibinfo{author}{G.~F{\'a}th},
\newblock \bibinfo{title}{Evolutionary games on graphs},
\newblock \bibinfo{journal}{Phys. Rep.} \bibinfo{volume}{446}
  (\bibinfo{year}{2007}) \bibinfo{pages}{97--216}.
\bibitem[{Roca et~al.(2009)Roca, Cuesta, and S{\'a}nchez}]{roca_plr09}
\bibinfo{author}{C.~P. Roca}, \bibinfo{author}{J.~A. Cuesta},
  \bibinfo{author}{A.~S{\'a}nchez},
\newblock \bibinfo{title}{Evolutionary game theory: Temporal and spatial
  effects beyond replicator dynamics},
\newblock \bibinfo{journal}{Phys. Life Rev.} \bibinfo{volume}{6}
  (\bibinfo{year}{2009}) \bibinfo{pages}{208--249}.
\bibitem[{Szab{\'o} and T{\H{o}}ke(1998)}]{szabo1998evolutionary}
\bibinfo{author}{G.~Szab{\'o}}, \bibinfo{author}{C.~T{\H{o}}ke},
\newblock \bibinfo{title}{Evolutionary prisoner’s dilemma game on a square
  lattice},
\newblock \bibinfo{journal}{Phys. Rev. E} \bibinfo{volume}{58}
  (\bibinfo{year}{1998}) \bibinfo{pages}{69}.
\bibitem[{Nowak and May(1992)}]{nowak1992evolutionary}
\bibinfo{author}{M.~A. Nowak}, \bibinfo{author}{R.~M. May},
\newblock \bibinfo{title}{Evolutionary games and spatial chaos},
\newblock \bibinfo{journal}{Nature} \bibinfo{volume}{359}
  (\bibinfo{year}{1992}) \bibinfo{pages}{826--829}.
\bibitem[{Ohtsuki and Nowak(2006)}]{ohtsuki_jtb06}
\bibinfo{author}{H.~Ohtsuki}, \bibinfo{author}{M.~A. Nowak},
\newblock \bibinfo{title}{The replicator equation on graphs},
\newblock \bibinfo{journal}{J. Theor. Biol.} \bibinfo{volume}{243}
  (\bibinfo{year}{2006}) \bibinfo{pages}{86--97}.
\bibitem[{Allen et~al.(2017)Allen, Lippner, Chen, Fotouhi, Momeni, Yau, and
  Nowak}]{allen2017evolutionary}
\bibinfo{author}{B.~Allen}, \bibinfo{author}{G.~Lippner},
  \bibinfo{author}{Y.-T. Chen}, \bibinfo{author}{B.~Fotouhi},
  \bibinfo{author}{N.~Momeni}, \bibinfo{author}{S.-T. Yau},
  \bibinfo{author}{M.~A. Nowak},
\newblock \bibinfo{title}{Evolutionary dynamics on any population structure},
\newblock \bibinfo{journal}{Nature} \bibinfo{volume}{544}
  (\bibinfo{year}{2017}) \bibinfo{pages}{227--230}.
\bibitem[{Traulsen et~al.(2007)Traulsen, Pacheco, and Nowak}]{traulsen_jtb07b}
\bibinfo{author}{A.~Traulsen}, \bibinfo{author}{J.~M. Pacheco},
  \bibinfo{author}{M.~A. Nowak},
\newblock \bibinfo{title}{Pairwise comparison and selection temperature in
  evolutionary game dynamics},
\newblock \bibinfo{journal}{J. Theor. Biol.} \bibinfo{volume}{246}
  (\bibinfo{year}{2007}) \bibinfo{pages}{522--529}.
\bibitem[{Fu et~al.(2009)Fu, Wang, Nowak, and Hauert}]{fu_pre09b}
\bibinfo{author}{F.~Fu}, \bibinfo{author}{L.~Wang}, \bibinfo{author}{M.~A.
  Nowak}, \bibinfo{author}{C.~Hauert},
\newblock \bibinfo{title}{Evolutionary dynamics on graphs: Efficient method for
  weak selection},
\newblock \bibinfo{journal}{Phys. Rev. E} \bibinfo{volume}{79}
  (\bibinfo{year}{2009}) \bibinfo{pages}{046707}.
\bibitem[{Zhou et~al.(2018)Zhou, Wu, Vasconcelos, and Wang}]{zhou_l_pre18}
\bibinfo{author}{L.~Zhou}, \bibinfo{author}{B.~Wu}, \bibinfo{author}{V.~V.
  Vasconcelos}, \bibinfo{author}{L.~Wang},
\newblock \bibinfo{title}{Simple property of heterogeneous aspiration dynamics:
  Beyond weak selection},
\newblock \bibinfo{journal}{Phys. Rev. E} \bibinfo{volume}{98}
  (\bibinfo{year}{2018}) \bibinfo{pages}{062124}.
\bibitem[{Lieberman et~al.(2005)Lieberman, Hauert, and
  Nowak}]{lieberman2005evolutionary}
\bibinfo{author}{E.~Lieberman}, \bibinfo{author}{C.~Hauert},
  \bibinfo{author}{M.~A. Nowak},
\newblock \bibinfo{title}{Evolutionary dynamics on graphs},
\newblock \bibinfo{journal}{Nature} \bibinfo{volume}{433}
  (\bibinfo{year}{2005}) \bibinfo{pages}{312--316}.
\bibitem[{Ohtsuki et~al.(2006)Ohtsuki, Hauert, Lieberman, and
  Nowak}]{ohtsuki2006simple}
\bibinfo{author}{H.~Ohtsuki}, \bibinfo{author}{C.~Hauert},
  \bibinfo{author}{E.~Lieberman}, \bibinfo{author}{M.~A. Nowak},
\newblock \bibinfo{title}{A simple rule for the evolution of cooperation on
  graphs and social networks},
\newblock \bibinfo{journal}{Nature} \bibinfo{volume}{441}
  (\bibinfo{year}{2006}) \bibinfo{pages}{502--505}.
\bibitem[{Ibsen-Jensen et~al.(2015)Ibsen-Jensen, Chatterjee, and
  Nowak}]{ibsen2015computational}
\bibinfo{author}{R.~Ibsen-Jensen}, \bibinfo{author}{K.~Chatterjee},
  \bibinfo{author}{M.~A. Nowak},
\newblock \bibinfo{title}{Computational complexity of ecological and
  evolutionary spatial dynamics},
\newblock \bibinfo{journal}{Proc. Natl. Acad. Sci. U.S.A.}
  \bibinfo{volume}{112} (\bibinfo{year}{2015}) \bibinfo{pages}{15636--15641}.
\bibitem[{Liu et~al.(2019)Liu, Jia, and Rong}]{liu_rr_amc19}
\bibinfo{author}{R.-R. Liu}, \bibinfo{author}{C.-X. Jia},
  \bibinfo{author}{Z.~Rong},
\newblock \bibinfo{title}{Effects of enhancement level on evolutionary public
  goods game with payoff aspirations},
\newblock \bibinfo{journal}{Appl. Math. Comput.} \bibinfo{volume}{350}
  (\bibinfo{year}{2019}) \bibinfo{pages}{242--248}.
\bibitem[{Wang et~al.(2013)Wang, Szolnoki, and Perc}]{wang2013interdependent}
\bibinfo{author}{Z.~Wang}, \bibinfo{author}{A.~Szolnoki},
  \bibinfo{author}{M.~Perc},
\newblock \bibinfo{title}{Interdependent network reciprocity in evolutionary
  games},
\newblock \bibinfo{journal}{Sci. Rep.} \bibinfo{volume}{3}
  (\bibinfo{year}{2013}) \bibinfo{pages}{1183}.
\bibitem[{Quan et~al.(2019)Quan, Li, and Wang}]{quan_j_c19}
\bibinfo{author}{J.~Quan}, \bibinfo{author}{X.~Li}, \bibinfo{author}{X.~Wang},
\newblock \bibinfo{title}{The evolution of cooperation in spatial public goods
  game with conditional peer exclusion},
\newblock \bibinfo{journal}{Chaos} \bibinfo{volume}{29} (\bibinfo{year}{2019})
  \bibinfo{pages}{103137}.
\bibitem[{Zhang et~al.(2021)Zhang, Huang, Li, Dai, and Yang}]{zhang_lm_pa21}
\bibinfo{author}{L.~Zhang}, \bibinfo{author}{C.~Huang},
  \bibinfo{author}{H.~Li}, \bibinfo{author}{Q.~Dai}, \bibinfo{author}{J.~Yang},
\newblock \bibinfo{title}{Cooperation guided by imitation, aspiration and
  conformity-driven dynamics in evolutionary games},
\newblock \bibinfo{journal}{Physica A} \bibinfo{volume}{561}
  (\bibinfo{year}{2021}) \bibinfo{pages}{125260}.
\bibitem[{Li et~al.(2021)Li, Mao, Wei, and Cong}]{li_k_csf21}
\bibinfo{author}{K.~Li}, \bibinfo{author}{Y.~Mao}, \bibinfo{author}{Z.~Wei},
  \bibinfo{author}{R.~Cong},
\newblock \bibinfo{title}{Pool-rewarding in n-person snowdrift game},
\newblock \bibinfo{journal}{Chaos, Solit. and Fract.} \bibinfo{volume}{143}
  (\bibinfo{year}{2021}) \bibinfo{pages}{110591}.
\bibitem[{Liu and Chen(2022)}]{liu_lj_rspa22}
\bibinfo{author}{L.~Liu}, \bibinfo{author}{X.~Chen},
\newblock \bibinfo{title}{Indirect exclusion can promote cooperation in
  repeated group interactions},
\newblock \bibinfo{journal}{Proc. R. Soc. A} \bibinfo{volume}{478}
  (\bibinfo{year}{2022}) \bibinfo{pages}{20220290}.
\bibitem[{Ohdaira(2022)}]{ohdaira_srep22}
\bibinfo{author}{T.~Ohdaira},
\newblock \bibinfo{title}{The probabilistic pool punishment proportional to the
  difference of payoff outperforms previous pool and peer punishment},
\newblock \bibinfo{journal}{Sci. Rep.} \bibinfo{volume}{12}
  (\bibinfo{year}{2022}) \bibinfo{pages}{6604}.
\bibitem[{Han(2022)}]{han_jrsif22}
\bibinfo{author}{T.~A. Han},
\newblock \bibinfo{title}{Institutional incentives for the evolution of
  committed cooperation: ensuring participation is as important as enhancing
  compliance},
\newblock \bibinfo{journal}{J. R. Soc. Interface} \bibinfo{volume}{19}
  (\bibinfo{year}{2022}) \bibinfo{pages}{20220036}.
\bibitem[{Nowak(2006)}]{nowak2006five}
\bibinfo{author}{M.~A. Nowak},
\newblock \bibinfo{title}{Five rules for the evolution of cooperation},
\newblock \bibinfo{journal}{Science} \bibinfo{volume}{314}
  (\bibinfo{year}{2006}) \bibinfo{pages}{1560--1563}.
\bibitem[{Perc et~al.(2017)Perc, Jordan, Rand, Wang, Boccaletti, and
  Szolnoki}]{perc2017statistical}
\bibinfo{author}{M.~Perc}, \bibinfo{author}{J.~J. Jordan},
  \bibinfo{author}{D.~G. Rand}, \bibinfo{author}{Z.~Wang},
  \bibinfo{author}{S.~Boccaletti}, \bibinfo{author}{A.~Szolnoki},
\newblock \bibinfo{title}{Statistical physics of human cooperation},
\newblock \bibinfo{journal}{Phys. Rep.} \bibinfo{volume}{687}
  (\bibinfo{year}{2017}) \bibinfo{pages}{1--51}.
\bibitem[{Rand and Nowak(2013)}]{rand_tcs13}
\bibinfo{author}{D.~A. Rand}, \bibinfo{author}{M.~A. Nowak},
\newblock \bibinfo{title}{Human cooperation},
\newblock \bibinfo{journal}{Trends in Cognitive Sciences} \bibinfo{volume}{17}
  (\bibinfo{year}{2013}) \bibinfo{pages}{413--425}.
\bibitem[{Perc et~al.(2013)Perc, G{\'o}mez-Garde{\~n}es, Szolnoki, and
  Flor{\'{\i}a and Y. Moreno}}]{perc_jrsi13}
\bibinfo{author}{M.~Perc}, \bibinfo{author}{J.~G{\'o}mez-Garde{\~n}es},
  \bibinfo{author}{A.~Szolnoki}, \bibinfo{author}{L.~M. Flor{\'{\i}a and Y.
  Moreno}},
\newblock \bibinfo{title}{Evolutionary dynamics of group interactions on
  structured populations: a review},
\newblock \bibinfo{journal}{J. R. Soc. Interface} \bibinfo{volume}{10}
  (\bibinfo{year}{2013}) \bibinfo{pages}{20120997}.
\bibitem[{Amaral and Javarone(2020)}]{amaral_rspa20}
\bibinfo{author}{M.~A. Amaral}, \bibinfo{author}{M.~A. Javarone},
\newblock \bibinfo{title}{Heterogeneity in evolutionary games: an analysis of
  the risk perception},
\newblock \bibinfo{journal}{Proc. R. Soc. A} \bibinfo{volume}{476}
  (\bibinfo{year}{2020}) \bibinfo{pages}{20200116}.
\bibitem[{Wang et~al.(2014)Wang, Szolnoki, and Perc}]{wang_z_pre14b}
\bibinfo{author}{Z.~Wang}, \bibinfo{author}{A.~Szolnoki},
  \bibinfo{author}{M.~Perc},
\newblock \bibinfo{title}{Different perceptions of social dilemmas:
  Evolutionary multigames in structured populations},
\newblock \bibinfo{journal}{Phys. Rev. E} \bibinfo{volume}{90}
  (\bibinfo{year}{2014}) \bibinfo{pages}{032813}.
\bibitem[{Huang et~al.(2018)Huang, Liu, Zhang, Yang, and Wang}]{huang_k_pa18}
\bibinfo{author}{K.~Huang}, \bibinfo{author}{Y.~Liu},
  \bibinfo{author}{Y.~Zhang}, \bibinfo{author}{C.~Yang},
  \bibinfo{author}{Z.~Wang},
\newblock \bibinfo{title}{Understanding cooperative behavior of agents with
  heterogeneous perceptions in dynamic networks},
\newblock \bibinfo{journal}{Physica A} \bibinfo{volume}{509}
  (\bibinfo{year}{2018}) \bibinfo{pages}{234--240}.
\bibitem[{Li et~al.(2016)Li, Szolnoki, Cong, and Wang}]{li_k_srep16}
\bibinfo{author}{K.~Li}, \bibinfo{author}{A.~Szolnoki},
  \bibinfo{author}{R.~Cong}, \bibinfo{author}{L.~Wang},
\newblock \bibinfo{title}{The coevolution of overconfidence and bluffing in the
  resource competition game},
\newblock \bibinfo{journal}{Sci. Rep.} \bibinfo{volume}{6}
  (\bibinfo{year}{2016}) \bibinfo{pages}{21104}.
\bibitem[{Szolnoki and Chen(2018)}]{szolnoki_pre18}
\bibinfo{author}{A.~Szolnoki}, \bibinfo{author}{X.~Chen},
\newblock \bibinfo{title}{Reciprocity-based cooperative phalanx maintained by
  overconfident players},
\newblock \bibinfo{journal}{Phys. Rev. E} \bibinfo{volume}{98}
  (\bibinfo{year}{2018}) \bibinfo{pages}{022309}.
\bibitem[{Szolnoki et~al.(2009)Szolnoki, Perc, Szab{\'o}, and
  Stark}]{szolnoki2009impact}
\bibinfo{author}{A.~Szolnoki}, \bibinfo{author}{M.~Perc},
  \bibinfo{author}{G.~Szab{\'o}}, \bibinfo{author}{H.-U. Stark},
\newblock \bibinfo{title}{Impact of aging on the evolution of cooperation in
  the spatial prisoner’s dilemma game},
\newblock \bibinfo{journal}{Phys. Rev. E} \bibinfo{volume}{80}
  (\bibinfo{year}{2009}) \bibinfo{pages}{021901}.
\bibitem[{Liu et~al.(2010)Liu, Rong, Jia, and Wang}]{liu2010effects}
\bibinfo{author}{R.-R. Liu}, \bibinfo{author}{Z.~Rong}, \bibinfo{author}{C.-X.
  Jia}, \bibinfo{author}{B.-H. Wang},
\newblock \bibinfo{title}{Effects of diverse inertia on scale-free--networked
  prisoner's dilemma games},
\newblock \bibinfo{journal}{EPL} \bibinfo{volume}{91} (\bibinfo{year}{2010})
  \bibinfo{pages}{20002}.
\bibitem[{Zhang et~al.(2011)Zhang, Fu, Wu, Xie, and Wang}]{zhang2011inertia}
\bibinfo{author}{Y.~Zhang}, \bibinfo{author}{F.~Fu}, \bibinfo{author}{T.~Wu},
  \bibinfo{author}{G.~Xie}, \bibinfo{author}{L.~Wang},
\newblock \bibinfo{title}{Inertia in strategy switching transforms the strategy
  evolution},
\newblock \bibinfo{journal}{Phys. Rev. E} \bibinfo{volume}{84}
  (\bibinfo{year}{2011}) \bibinfo{pages}{066103}.
\bibitem[{Szab{\'o} and Hauert(2002)}]{szabo2002phase}
\bibinfo{author}{G.~Szab{\'o}}, \bibinfo{author}{C.~Hauert},
\newblock \bibinfo{title}{Phase transitions and volunteering in spatial public
  goods games},
\newblock \bibinfo{journal}{Phys. Rev. Lett.} \bibinfo{volume}{89}
  (\bibinfo{year}{2002}) \bibinfo{pages}{118101}.
\bibitem[{Du et~al.(2012)Du, Cao, Liu, and Wang}]{du2012effects}
\bibinfo{author}{W.-B. Du}, \bibinfo{author}{X.-B. Cao}, \bibinfo{author}{R.-R.
  Liu}, \bibinfo{author}{Z.~Wang},
\newblock \bibinfo{title}{Effects of inertia on evolutionary prisoner's dilemma
  game},
\newblock \bibinfo{journal}{Commun. Theor. Phys.} \bibinfo{volume}{58}
  (\bibinfo{year}{2012}) \bibinfo{pages}{451}.
\bibitem[{Chang et~al.(2018)Chang, Zhang, Wu, and Xie}]{chang2018cooperation}
\bibinfo{author}{S.~Chang}, \bibinfo{author}{Z.~Zhang},
  \bibinfo{author}{Y.~Wu}, \bibinfo{author}{Y.~Xie},
\newblock \bibinfo{title}{Cooperation is enhanced by inhomogeneous inertia in
  spatial prisoner’s dilemma game},
\newblock \bibinfo{journal}{Physica A} \bibinfo{volume}{490}
  (\bibinfo{year}{2018}) \bibinfo{pages}{419--425}.
\bibitem[{Wang and Szolnoki(2023)}]{wang2023evolution}
\bibinfo{author}{C.~Wang}, \bibinfo{author}{A.~Szolnoki},
\newblock \bibinfo{title}{Evolution of cooperation under a generalized
  death-birth process},
\newblock \bibinfo{journal}{Physical Review E} \bibinfo{volume}{107}
  (\bibinfo{year}{2023}) \bibinfo{pages}{024303}.
\bibitem[{Jia et~al.(2018)Jia, Jin, Du, and Shi}]{jia2018effects}
\bibinfo{author}{D.~Jia}, \bibinfo{author}{J.~Jin}, \bibinfo{author}{C.~Du},
  \bibinfo{author}{L.~Shi},
\newblock \bibinfo{title}{Effects of inertia on the evolution of cooperation in
  the voluntary prisoner’s dilemma game},
\newblock \bibinfo{journal}{Physica A} \bibinfo{volume}{509}
  (\bibinfo{year}{2018}) \bibinfo{pages}{817--826}.
\bibitem[{He et~al.(2020)He, Wang, and Yu}]{he2020behavior}
\bibinfo{author}{J.~He}, \bibinfo{author}{J.~Wang}, \bibinfo{author}{F.~Yu},
\newblock \bibinfo{title}{Behavior inertia of individuals promotes cooperation
  in spatial prisoner's dilemma game},
\newblock \bibinfo{journal}{EPL} \bibinfo{volume}{132} (\bibinfo{year}{2020})
  \bibinfo{pages}{38002}.
\bibitem[{Allen and Nowak(2014)}]{allen2014games}
\bibinfo{author}{B.~Allen}, \bibinfo{author}{M.~A. Nowak},
\newblock \bibinfo{title}{Games on graphs},
\newblock \bibinfo{journal}{EMS Surv. Math. Sci.} \bibinfo{volume}{1}
  (\bibinfo{year}{2014}) \bibinfo{pages}{113--151}.
\bibitem[{Su et~al.(2019)Su, Li, Wang, and Eugene~Stanley}]{su2019spatial}
\bibinfo{author}{Q.~Su}, \bibinfo{author}{A.~Li}, \bibinfo{author}{L.~Wang},
  \bibinfo{author}{H.~Eugene~Stanley},
\newblock \bibinfo{title}{Spatial reciprocity in the evolution of cooperation},
\newblock \bibinfo{journal}{Proc. R. Soc. B} \bibinfo{volume}{286}
  (\bibinfo{year}{2019}) \bibinfo{pages}{20190041}.
\bibitem[{Su et~al.(2018)Su, Wang, and Stanley}]{su2018understanding}
\bibinfo{author}{Q.~Su}, \bibinfo{author}{L.~Wang}, \bibinfo{author}{H.~E.
  Stanley},
\newblock \bibinfo{title}{Understanding spatial public goods games on
  three-layer networks},
\newblock \bibinfo{journal}{New J. Phys.} \bibinfo{volume}{20}
  (\bibinfo{year}{2018}) \bibinfo{pages}{103030}.
\bibitem[{Li et~al.(2014)Li, Wu, and Wang}]{li2014cooperation}
\bibinfo{author}{A.~Li}, \bibinfo{author}{B.~Wu}, \bibinfo{author}{L.~Wang},
\newblock \bibinfo{title}{Cooperation with both synergistic and local
  interactions can be worse than each alone},
\newblock \bibinfo{journal}{Sci. Rep.} \bibinfo{volume}{4}
  (\bibinfo{year}{2014}) \bibinfo{pages}{5536}.
\bibitem[{Li and Wang(2015)}]{li2015evolutionary}
\bibinfo{author}{A.~Li}, \bibinfo{author}{L.~Wang},
\newblock \bibinfo{title}{Evolutionary dynamics of synergistic and discounted
  group interactions in structured populations},
\newblock \bibinfo{journal}{J. Theor. Biol.} \bibinfo{volume}{377}
  (\bibinfo{year}{2015}) \bibinfo{pages}{57--65}.
\bibitem[{Li et~al.(2016)Li, Broom, Du, and Wang}]{li2016evolutionary}
\bibinfo{author}{A.~Li}, \bibinfo{author}{M.~Broom}, \bibinfo{author}{J.~Du},
  \bibinfo{author}{L.~Wang},
\newblock \bibinfo{title}{Evolutionary dynamics of general group interactions
  in structured populations},
\newblock \bibinfo{journal}{Phys. Rev. E} \bibinfo{volume}{93}
  (\bibinfo{year}{2016}) \bibinfo{pages}{022407}.
\bibitem[{Perc et~al.(2013)Perc, G{\'o}mez-Gardenes, Szolnoki, Flor{\'\i}a, and
  Moreno}]{perc2013evolutionary}
\bibinfo{author}{M.~Perc}, \bibinfo{author}{J.~G{\'o}mez-Gardenes},
  \bibinfo{author}{A.~Szolnoki}, \bibinfo{author}{L.~M. Flor{\'\i}a},
  \bibinfo{author}{Y.~Moreno},
\newblock \bibinfo{title}{Evolutionary dynamics of group interactions on
  structured populations: a review},
\newblock \bibinfo{journal}{J. R. Soc. Interface} \bibinfo{volume}{10}
  (\bibinfo{year}{2013}) \bibinfo{pages}{20120997}.
\bibitem[{Battiston et~al.(2020)Battiston, Cencetti, Iacopini, Latora, Lucas,
  Patania, Young, and Petri}]{battiston2020networks}
\bibinfo{author}{F.~Battiston}, \bibinfo{author}{G.~Cencetti},
  \bibinfo{author}{I.~Iacopini}, \bibinfo{author}{V.~Latora},
  \bibinfo{author}{M.~Lucas}, \bibinfo{author}{A.~Patania},
  \bibinfo{author}{J.-G. Young}, \bibinfo{author}{G.~Petri},
\newblock \bibinfo{title}{Networks beyond pairwise interactions: structure and
  dynamics},
\newblock \bibinfo{journal}{Phys. Rep.} \bibinfo{volume}{874}
  (\bibinfo{year}{2020}) \bibinfo{pages}{1--92}.
\bibitem[{Burgio et~al.(2020)Burgio, Matamalas, G{\'o}mez, and
  Arenas}]{burgio2020evolution}
\bibinfo{author}{G.~Burgio}, \bibinfo{author}{J.~T. Matamalas},
  \bibinfo{author}{S.~G{\'o}mez}, \bibinfo{author}{A.~Arenas},
\newblock \bibinfo{title}{Evolution of cooperation in the presence of
  higher-order interactions: from networks to hypergraphs},
\newblock \bibinfo{journal}{Entropy} \bibinfo{volume}{22}
  (\bibinfo{year}{2020}) \bibinfo{pages}{744}.
\bibitem[{Alvarez-Rodriguez et~al.(2021)Alvarez-Rodriguez, Battiston,
  de~Arruda, Moreno, Perc, and Latora}]{alvarez2021evolutionary}
\bibinfo{author}{U.~Alvarez-Rodriguez}, \bibinfo{author}{F.~Battiston},
  \bibinfo{author}{G.~F. de~Arruda}, \bibinfo{author}{Y.~Moreno},
  \bibinfo{author}{M.~Perc}, \bibinfo{author}{V.~Latora},
\newblock \bibinfo{title}{Evolutionary dynamics of higher-order interactions in
  social networks},
\newblock \bibinfo{journal}{Nature Human Behav.} \bibinfo{volume}{5}
  (\bibinfo{year}{2021}) \bibinfo{pages}{586--595}.
\bibitem[{Cox and Griffeath(1983)}]{cox_ap83}
\bibinfo{author}{J.~T. Cox}, \bibinfo{author}{D.~Griffeath},
\newblock \bibinfo{title}{Occupation time limit theorems for the voter model},
\newblock \bibinfo{journal}{Ann. Probab.} \bibinfo{volume}{11}
  (\bibinfo{year}{1983}) \bibinfo{pages}{876--893}.
\bibitem[{Nowak et~al.(2004)Nowak, Sasaki, Taylor, and
  Fudenberg}]{nowak2004emergence}
\bibinfo{author}{M.~A. Nowak}, \bibinfo{author}{A.~Sasaki},
  \bibinfo{author}{C.~Taylor}, \bibinfo{author}{D.~Fudenberg},
\newblock \bibinfo{title}{Emergence of cooperation and evolutionary stability
  in finite populations},
\newblock \bibinfo{journal}{Nature} \bibinfo{volume}{428}
  (\bibinfo{year}{2004}) \bibinfo{pages}{646--650}.
\bibitem[{Nowak et~al.(2010)Nowak, Tarnita, and Wilson}]{nowak2010evolution}
\bibinfo{author}{M.~A. Nowak}, \bibinfo{author}{C.~E. Tarnita},
  \bibinfo{author}{E.~O. Wilson},
\newblock \bibinfo{title}{The evolution of eusociality},
\newblock \bibinfo{journal}{Nature} \bibinfo{volume}{466}
  (\bibinfo{year}{2010}) \bibinfo{pages}{1057--1062}.
\bibitem[{Dornic et~al.(2001)Dornic, Chat{\'e}, Chave, and
  Hinrichsen}]{dornic_prl01}
\bibinfo{author}{I.~Dornic}, \bibinfo{author}{H.~Chat{\'e}},
  \bibinfo{author}{J.~Chave}, \bibinfo{author}{H.~Hinrichsen},
\newblock \bibinfo{title}{Critical coarsening without surface tension: The
  universality class of the voter model},
\newblock \bibinfo{journal}{Phys. Rev. Lett.} \bibinfo{volume}{87}
  (\bibinfo{year}{2001}) \bibinfo{pages}{045701}.
\bibitem[{Liu et~al.(2010)Liu, Rong, Jia, and Wang}]{liu_rr_epl10}
\bibinfo{author}{R.-R. Liu}, \bibinfo{author}{Z.~Rong}, \bibinfo{author}{C.-X.
  Jia}, \bibinfo{author}{B.-H. Wang},
\newblock \bibinfo{title}{Effects of diverse inertia on scale-free-networked
  prisoner's dilemma games},
\newblock \bibinfo{journal}{EPL} \bibinfo{volume}{91} (\bibinfo{year}{2010})
  \bibinfo{pages}{20002}.

\end{thebibliography}


\end{document}